\documentclass[preprint,review,12pt]{elsarticle}

\usepackage{amssymb}
\usepackage{amsthm}
\usepackage[fleqn]{amsmath}
\usepackage{color,soul}
\usepackage{floatrow}
\usepackage{graphicx}
\usepackage{caption}
\usepackage{subcaption}
\usepackage{siunitx}
\usepackage{nicefrac}
\usepackage[label font=bf,labelformat=simple]{subfig}
\usepackage[font=small,skip=0pt]{caption}
\usepackage{epsfig}
\usepackage[version=4]{mhchem}
\usepackage[symbol]{footmisc}

\usepackage[dvipsnames]{xcolor}
\usepackage{ulem}
\usepackage{enumitem}
\usepackage{multirow}
\usepackage{anyfontsize}

\biboptions{sort&compress}

\usepackage{lineno}

\begin{document}

\begin{frontmatter}

\title{JPResUnet: A Joint Probability Density Function Translation Model in Partially Premixed Flames}

\author[1]{Hanying~Yang}

\author[1,2]{James~C.~Massey}
\author[1]{Nedunchezhian~Swaminathan}

\address[1]{Department~of~Engineering, University~of~Cambridge, Trumpington~Street, Cambridge~CB2~1PZ,~United~Kingdom}
\address[2]{Robinson~College, University~of~Cambridge, Grange~Road, Cambridge~CB3~9AN,~United~Kingdom}

\begin{abstract}
Machine learning (ML) models are often constrained by their limitations in extrapolation, which restricts their applicability in engineering contexts. Conversely, while exhibiting broad generality, many established scientific models seem to lack the necessary accuracy. This study addresses these challenges by introducing JPResUnet (Joint PDF Residual U-net), a novel model that integrates the strengths of both ML and traditional scientific approaches to predict sub-grid joint probability density functions (PDFs) in partially premixed flames. JPResUnet employs a residual U-Net architecture to translate classic $\beta$-PDFs to sub-grid PDFs. The model is trained using direct numerical simulation (DNS) data from methane--air moderate or intense low-oxygen dilution (MILD) combustion and is initially tested through \textit{a priori} assessments on out-of-sample data. Comparative analyses against an artificial neural network (ANN) and the $\beta$-PDF approach demonstrate that JPResUnet consistently outperforms these methods in capturing complex sub-grid features with greater accuracy and robustness for both box and Gaussian kernels of varying widths. Further evaluations on a different case reveal the model’s generalisability, where the ANN is unable to produce satisfactory prediction. Subsequent \textit{a posteriori} assessment involves two versions of JPResUnet with different output PDF resolutions, which are deployed for large eddy simulation (LES) of a multi-regime burner through the look-up table (LUT) approach. The higher resolution model yields improvements in temperature estimates compared to the conventional LUT method. This highlights the potential of the JPResUnet model for robust and accurate LES of reacting flows with ML.
\end{abstract}

\begin{keyword}
Joint probability density function \sep Partially premixed flame \sep Residual U-net \sep MILD combustion \sep Multi-regime combustion
\end{keyword}
\end{frontmatter}

\newpage
\section{Introduction} \label{sec:intro}
Large eddy simulation (LES) has emerged as an important tool for the simulation of large-scale combustion systems, particularly in scenarios where fully resolving flow fields is computationally prohibitive. By employing a spatial filter, LES effectively resolves physical processes associated with scales larger than the filter width $\Delta$, while smaller scales, collectively referred to as the sub-grid scale (SGS) range, are modelled. Over the past half-century, turbulent convection and diffusion within the SGS range have been extensively studied, leading to the development of accurate models for both non-reacting \cite{Pope2000} and reacting flows \cite{Poinsot2005}. However, modelling sub-grid reactions presents significant challenges due to the highly nonlinear nature of chemical processes and their complex interactions with turbulence \cite{Poinsot2005}. Although presumed probability density functions (PDFs) have been widely utilised to capture sub-grid fluctuations in thermo-chemical quantities for premixed \cite{Vreman2009,Hernandez2011}, non-premixed \cite{Cook1994,Ihme2008a,Ihme2008b}, and partially premixed \cite{Bradley1991,Donini2017,Domingo2002,Michel2009} flames, this approach can encounter limitations under specific conditions, such as combustion with multiple fuel streams of varying composition, multi-regime combustion, and multi-phase flows. In the context of partially premixed flames, additional difficulties arise due to the strong correlation between the progress variable and the mixture fraction, whereas the presumed joint PDF approach typically assumes statistical independence between these two scalars at the sub-grid level \cite{Chen2017,See2015}. 

The use of machine learning (ML) as an alternative to traditional reaction rate modelling has been explored for over 30 years, with early studies focusing on the development of artificial neural networks (ANNs) to model chemical kinetics \cite{Christo1995,Christo1996a,Christo1996b,Blasco1998,Blasco1999,Sen2009,Sen2010a,Sen2010b,Chatzopoulos2013}. Ihme \textit{et al.} \cite{Ihme2008,Ihme2009} later proposed a systematic approach to optimise these techniques. The advent of deep learning has led to the exploration of more complex ANN structures for sub-grid PDF modelling. De Frahan \textit{et al.} \cite {T.HenrydeFrahan2019} validated the predictive capabilities of ANNs for the marginal PDF of the progress variable in swirling methane--air premixed flames. Similarly, Yao \textit{et al.} \cite{Yao2020} demonstrated that ANNs could predict the marginal PDF of the mixture fraction in turbulent spray flames more accurately than conventional presumed PDFs. Chen \textit{et al.} \cite{Chen2021} further extended this approach to predict the joint PDF of the progress variable and mixture fraction in a partially premixed flame, specifically in moderate or intense low-oxygen dilution (MILD) combustion. Despite these advancements, most studies are self-testing, meaning that the training and testing data share similar thermo-chemical and -physical conditions. To assess the generalisability of ANNs, Yang \textit{et al.} \cite{Yang2023} tested a model trained on MILD combustion data across five different premixed flames. The ANN performed well when applied to flames with the same fuel as in the MILD cases. However, its predictive accuracy diminished when there were deviations from the training conditions, such as increased filter size or a shift to hydrogen--air flames, underscoring the challenges for developing ML models with broad generalisability.

The rapid advancements in deep learning have led to the emergence of novel ML architectures with exceptional predictive capabilities, particularly in domains such as computer vision and natural language processing \cite{Brown2020,Dosovitskiy2020}. These advances have also been leveraged in many fields, including biology and material science \cite{Jumper2021,Merchant2023}. In the context of combustion, Xing \textit{et al.} \cite{Xing2021} employed a deep convolutional neural network (CNN) with a U-net architecture, commonly used in image processing \cite{Ronneberger2015,Zhou2018}, to predict the progress variable SGS variance. Trained on a planar flame in homogeneous isotropic turbulence and tested on a complex slot jet flame, the model demonstrated good agreement with direct numerical simulation (DNS), thus verifying its generalisability. Nista \textit{et al.} \cite{Nista2023,Nista2024} utilised a generative adversarial network (GAN), originally designed for computer vision \cite{Ledig2017}, to recover fully resolved flow fields from LES input. This GAN outperformed other models across various thermo-chemical and -physical test conditions.

Inspired by the image translation techniques in computer vision, this study aims to develop and test a U-net-based model, referred to as JPResUnet (Joint PDF Residual U-net), to infer joint sub-grid PDFs based on presumed distributions while maintaining generalisability. The model is trained on data from a single MILD combustion case and tested on unseen samples from different cases with varying filter sizes and types. Additionally, \textit{a posteriori} assessment through LES for a multi-regime burner (MRB) is conducted to evaluate the model's practical performance. 

The structure of the paper is as follows. Section \ref{sec:theory} provides a theoretical overview of turbulent combustion modelling in LES and the concept of PDF translation. Section \ref{sec:data_process} details the methodology for training the JPResUnet model and describes the training dataset. Section \ref{sec:priori} compares the joint PDFs predicted by the JPResUnet and corresponding values from DNS, alongside a discussion of the filtered reaction rate modelled via these PDFs. Results from \textit{a posteriori} assessment are discussed in Section \ref{sec:posteriori}, and conclusions are summarised in the final section.

\section{Theoretical background} \label{sec:theory}
\subsection{Governing equations} \label{subsec:govening_eqns}
In terms of the LES implemented by present work, the mass, momentum and total enthalpy are transported as 
\begin{equation}
  \frac{\partial\overline{\rho}}{\partial t} + \mathbf{\nabla}\cdot (\overline{\rho}\widetilde{\mathbf{U}}) = 0,
  \label{eq:mass_conservation}
\end{equation}
\begin{equation}
  \overline{\rho}\frac{\text{D}\widetilde{\mathbf{U}}}{\text{D}t}  = -\mathbf{\nabla}\overline{p}+\mathbf{\nabla}\cdot \left [\overline{\tau}-\left (\overline{\rho}\widetilde{\mathbf{U}\mathbf{U}} - \overline{\rho}\widetilde{\mathbf{U}}\widetilde{\mathbf{U}} \right ) \right],
  \label{eq:moment_conservation}
\end{equation}
\begin{equation}
  \overline{\rho}\frac{\text{D}\widetilde{h}}{\text{D}t}  = \mathbf{\nabla}\cdot \left [\overline{\rho\alpha}\mathbf{\nabla}\widetilde{h}-\left (\overline{\rho}\widetilde{\mathbf{U}h} - \overline{\rho}\widetilde{\mathbf{U}}\widetilde{h} \right ) \right].
  \label{eq:enthalpy_conservation}
\end{equation}
The symbols $\rho$, $\mathbf{U}$ and $h$ represent density, velocity vector and thermo-chemical enthalpy (sum of sensible and chemical enthalpies) respectively. Filtering and Favre-filtering operations are denoted by $\overline{\cdot}$ and $\widetilde{\cdot}$, respectively. The operator $\text{D}/\text{D}t=\partial/\partial t + \widetilde{\mathbf{U}}\cdot\mathbf{\nabla}$ is the material derivative. The shear stress $\tau$ in Eq. (\ref{eq:moment_conservation}) is calculated as $\overline{\tau}=2\mu[\widetilde{\mathbf{S}} - 1/3(\mathbf{\nabla}\cdot \widetilde{\mathbf{U}})\mathbf{I}]$, where $\mu$ is the molecular dynamic viscosity, $\widetilde{\mathbf{S}}$ is the strain rate tensor defined as $0.5[\mathbf{\nabla}\widetilde{\mathbf{U}}+(\mathbf{\nabla}\widetilde{\mathbf{U}})^\text{T}]$, and $\mathbf{I}$ is the identity matrix. The term within the parentheses in Eq. (\ref{eq:moment_conservation}) represents the residual stress tensor $\overline{\tau}^{R}$. It is unclosed and modelled as $\overline{\tau}^R=2(\overline{\rho}\widetilde{k}_{\text{sgs}}\mathbf{I})/3-2\overline{\rho}\nu_T[\widetilde{\mathbf{S}} - (\mathbf{\nabla}\cdot \widetilde{\mathbf{U}})\mathbf{I}/3]$, where $\widetilde{k}_{\text{sgs}}$ is the turbulent kinetic energy at the sub-grid scale and $\nu_T$ is the turbulent viscosity. The latter is modelled using the $\sigma-$model as $\nu_T=(C_{\sigma}\Delta)^2\mathcal{D}_{\sigma}$ \cite{Nicoud2011}, where the model constant $C_{\sigma}=1.5$ is decided similar to \cite{Massey2023b}. The operator $\mathcal{D}_{\sigma}$ denotes the differential operator related to the singular values of the Jacobian matrix of the velocity. The molecular thermal diffusivity $\alpha$ in Eq. (\ref{eq:enthalpy_conservation}) is calculated as $\overline{\rho \alpha}=\overline{\rho}\nu/\text{Pr}$, where $\nu$ is the molecular kinematic viscosity, and $\text{Pr}=0.7$ is the Prandtl number. The sub-grid scalar flux in Eq. (\ref{eq:enthalpy_conservation}) is modelled as $\overline{\rho}\widetilde{\mathbf{U}h}-\overline{\rho}\widetilde{\mathbf{U}}\widetilde{h}=-\overline{\rho}(\nu_T/\text{Pr}_T)\mathbf{\nabla}\widetilde{h}$, where $\text{Pr}_T=0.7$ is the turbulent Prandtl number following the previous study on the MRB considered in the \textit{a posteriori} assessment \cite{Massey2023b}.

\subsection{Combustion closure} \label{subsec:combustion_eqns}
A joint PDF-based tabulation method \cite{Langella2016,Chen2017,Massey2021,Massey2022} is used for combustion modelling. During the simulation, unclosed thermo-chemical scalars are retrieved from a four-dimensional look-up table (LUT) parameterised by the filtered progress variable $\widetilde{c}$, the mixture fraction $\widetilde{Z}$ and their respective sub-grid variance $\widetilde{\sigma}_{c,\text{sgs}}^2$ and $\widetilde{\sigma}_{Z,\text{sgs}}^2$. The progress variable $c$ is defined as $(Y_{\text{CO}}+Y_{\text{CO}_2})/(Y_{\text{CO}}+Y_{\text{CO}_2})^{\text{eq}}$, where the superscript `eq' denotes the equilibrium state for a given mixture fraction following previous studies that used methane--air mixtures \cite{Fiorina2003,Chen2017}. The mixture fraction $Z$ is defined according to Bilger \cite{Bilger1989}. These quantities are transported using 
\begin{equation}
  \overline{\rho}\frac{\text{D}\widetilde{c}}{\text{D}t} \approx \mathbf{\nabla}\cdot \left (\overline{\rho \mathcal{D}}+\overline{\rho}\frac{\nu_T}{\text{Sc}_T} \right )\mathbf{\nabla}\widetilde{c} + \overline{\dot{\omega}^*},
  \label{eq:progress_variable_conservation}
\end{equation}
\begin{equation}
  \overline{\rho}\frac{\text{D}\widetilde{Z}}{\text{D}t} \approx \mathbf{\nabla}\cdot \left (\overline{\rho \mathcal{D}}+\overline{\rho}\frac{\nu_T}{\text{Sc}_T} \right )\mathbf{\nabla}\widetilde{Z},
  \label{eq:mixture_fraction_conservation}
\end{equation}
\begin{equation}
\begin{split}
  \overline{\rho}&\frac{\text{D}\widetilde{\sigma}_{c,\text{sgs}}^2}{\text{D}t} \approx \mathbf{\nabla}\cdot \left (\overline{\rho \mathcal{D}}+\overline{\rho}\frac{\nu_T}{\text{Sc}_T} \right )\mathbf{\nabla}\widetilde{\sigma}_{c,\text{sgs}}^2 - 2\overline{\rho}\widetilde{\chi}_{c,\text{sgs}}\\ 
  &+ 2\overline{\rho}\frac{\nu_T}{\text{Sc}_T}(\mathbf{\nabla}\widetilde{c} \cdot \mathbf{\nabla}\widetilde{c}) + 2(\overline{c\dot{\omega}^*}-\widetilde{c}\overline{\dot{\omega}^*}),
  \label{eq:progress_variable_variance_conservation}
\end{split}
\end{equation}
\begin{equation}
\begin{split}
  \overline{\rho}&\frac{\text{D}\widetilde{\sigma}_{Z,\text{sgs}}^2}{\text{D}t} \approx \mathbf{\nabla}\cdot \left (\overline{\rho \mathcal{D}}+\overline{\rho}\frac{\nu_T}{\text{Sc}_T} \right )\mathbf{\nabla}\widetilde{\sigma}_{Z,\text{sgs}}^2 - 2\overline{\rho}\widetilde{\chi}_{Z,\text{sgs}}\\ 
  &+ 2\overline{\rho}\frac{\nu_T}{\text{Sc}_T}(\mathbf{\nabla}\widetilde{Z} \cdot \mathbf{\nabla}\widetilde{Z}),
  \label{eq:mixture_fraction_variance_conservation}
\end{split}
\end{equation}
where the turbulent Schmidt number $\text{Sc}_T$ is assigned as a constant value of 0.4 following the previous study on the MRB considered in the \textit{a posteriori} assessment \cite{Massey2023b}. The term $\overline{\rho \mathcal{D}}$ is calculated as $\overline{\rho} \nu / \text{Sc}$ with the molecular Schmidt number $\text{Sc}=0.7$ \cite{Chen2020}. The sub-grid scalar dissipation rate (SDR) for the progress variable and mixture fraction are denoted respectively as $\widetilde{\chi}_{c,\text{sgs}}$ and $\widetilde{\chi}_{Z,\text{sgs}}$. $\widetilde{\chi}_{c,\text{sgs}}$ is modelled using a model proposed by Dunstan et al. \cite{Dunstan2013}, which has been validated by many studies \cite{Langella2016,Chen2017,Chen2020,Massey2021}. $\widetilde{\chi}_{Z,\text{sgs}}$ is modelled using a linear relaxation model as $\widetilde{\chi}_{Z,\text{sgs}}=C_Z(\nu_T/\Delta^2)\sigma_{Z,\text{sgs}}^2$ with a constant $C_Z=2$ \cite{Pierce1998,Pitsch2006,Mura2007}. 

The source term $\overline{\dot{\omega}^*}$ in Eq. (\ref{eq:progress_variable_conservation}) and (\ref{eq:progress_variable_variance_conservation}) originates from three components \cite{Domingo2002,Bray2005}: premixed combustion $\overline{\dot{\omega}}_{\text{p}}$, non-premixed combustion $\overline{\dot{\omega}}_{\text{np}}$, and their interaction through the cross dissipation rate $\overline{\dot{\omega}}_{\text{cdr}}$. This cross dissipation term is neglected following previous studies \cite{Bray2005,Ruan2014}. The premixed combustion term is modelled as
\begin{equation}
\overline{\dot{\omega}}_{\text{p}}=\overline{\rho} \int_{0}^{1} \int_{0}^{1} \frac{\dot{\omega}(\eta,\xi)}{\rho(\eta,\xi)}\widetilde{P}(\eta,\xi) \,\text{d}\eta \,\text{d}\xi,
\label{eq:premixed_reaction_rate}
\end{equation}
where $\eta$ and $\xi$ are the sample space variables for $c$ and $Z$ respectively. The flamelet reaction rate $\dot{\omega}$ and density $\rho$ are obtained by solving one-dimensional unstrained planar laminar premixed flames over the flammability of the methane--air flame using the GRI-Mech 3.0 chemical mechanism \cite{Smith} in Cantera \cite{Goodwin}. The non-premixed combustion mode is modelled by using the marginal PDF of the mixture fraction as 
\begin{equation}
\overline{\dot{\omega}}_{\text{np}}=\overline{\rho}\widetilde{c}\widetilde{\chi}_{Z} \int_{0}^{1} \frac{1}{\psi^{\text{eq}}(\xi)}\frac{d^2 \psi^{\text{eq}}(\xi)}{d \xi^2}\widetilde{P}(\xi) \,\text{d}\xi,
\label{eq:non_premixed_reaction_rate}
\end{equation}
where $\widetilde{\chi}_{Z}=\widetilde{D}(\mathbf{\nabla}\widetilde{Z} \cdot \mathbf{\nabla}\widetilde{Z})+\widetilde{\chi}_{Z,\text{sgs}}$, as seen in the work \cite{Chen2017} and $\psi=Y_{\text{CO}}+Y_{\text{CO}_2}$. Another source term in Eq. (\ref{eq:progress_variable_variance_conservation}) is assumed as $\overline{c\dot{\omega}^*}\approx \overline{c\dot{\omega}_\text{p}}$ \cite{Ruan2014}, and it is modelled similar to Eq. (\ref{eq:premixed_reaction_rate}). 

\subsection{Presumed joint PDF} \label{subect:joint_PDF}
Typically, the joint PDF in Eq. (\ref{eq:premixed_reaction_rate}) is calculated as the product of two marginal PDFs of progress variable and mixture fraction \cite{Massey2022,Massey2023b}, based on the assumption of the statistical independence between these two scalars at sub-grid level. Hence, 
\begin{equation}
\widetilde{P}(\eta,\xi)=\widetilde{P}_{\beta}(\eta;\widetilde{c},\widetilde{\sigma}_{c,\text{sgs}}^2) \times \widetilde{P}_{\beta}(\xi;\widetilde{Z},\widetilde{\sigma}_{Z,\text{sgs}}^2),
\label{eq:joint_PDF}
\end{equation}
where the marginal PDF is presumed with the $\beta$-PDF distribution. This presumed distribution is calculated as
\begin{equation}
  \widetilde{P}(\eta;\widetilde{c},\widetilde{\sigma}_{c,\text{sgs}}^2) = \frac{\Gamma(a+b)}{\Gamma(a)\Gamma(b)}\eta^{a-1}(1-\eta)^{b-1},
  \label{eq:beta_PDF}
\end{equation}
where $a=\widetilde{c}(1/\widetilde{g}_c-1)$, $b=(1-\widetilde{c})(1/\widetilde{g}_c-1)$ and $\Gamma$ is the $gamma$ function. $\widetilde{g}_c$ is the segregation factor, calculated as $\widetilde{g}_c=\widetilde{\sigma}_{c,\text{sgs}}^2/(\widetilde{c}(1-\widetilde{c}))$. The marginal PDF for the mixture fraction is calculated in a similar way. 

The statistical independence in Eq. (\ref{eq:joint_PDF}) may not hold in partially premixed flames due to the evident correlation between the progress variable and mixture fraction at the SGS level \cite{Chen2018,Chen2021} when the grid width is large. Alternative models, such as the \textit{copula} \cite{Plackett1965}, which accounts for the cross-correlation, have been suggested \cite{Chen2018} but require additional transport equations for sub-grid covariance between $c$ and $Z$, leading to increased complexity and need for further modelling. Consequently, the present work utilises the presumed joint PDF formulation (Eq. (\ref{eq:joint_PDF})) and considers incorporating \textit{copula} models in future studies. 

\subsection{PDF-to-PDF translation model} \label{subsec:PDF_translation}
\begin{sloppypar}
To improve the accuracy of PDF predictions, this study develops the JPResUnet model, a PDF-to-PDF translation framework inspired by supervised image-to-image translation techniques. This model translates from a source probability space $\{\Omega,\mathcal{F},\mathcal{P}\}_{A}$ to the target probability space $\{\Omega,\mathcal{F},\mathcal{P}\}_{B}$, where $\Omega$ is the sample space, $\mathcal{F}$ denotes the $\sigma$--algebra on $\Omega$, and $\mathcal{P}$ is the probability measure. Regarding the partially premixed flame, the sample space $\Omega$ is 2-dimensional, constituted by the progress variable $c \in [0,1]$ and mixture fraction $Z \in [0,1]$. The model is designed for self-translation, implying that the source and target probability spaces originate from the same flame field, with identical $\sigma-$algebras, i.e., $\mathcal{F}_{A}=\mathcal{F}_{B}=\mathcal{F}$. Therefore, the translation is constrained to the probability measure $\mathcal{P}$, which is described through the PDF $P$ for the continuous random variables. This process is displayed as,
\begin{equation}
    P_{AB}(\eta,\xi) = \mathcal{M}_{\theta}^{A \mapsto B}(P_A(\eta,\xi)) \text{, for } (\eta,\xi) \in \mathcal{F},
\label{eq:PDF_translation}
\end{equation}
where the subscript $\theta$ denotes the parameter of the model $\mathcal{M}$, and the outcome of the model, $P_{AB}$, is associated with the probability space $B$. 
\end{sloppypar}

The model is optimised by minimising the loss function $\mathcal{L}$ as 
\begin{equation}
\begin{split}
    \min_{\theta} \mathcal{L} &= \min_{\theta} \mathbb{E}_{\eta,\xi} \left [\parallel P_{AB}(\eta,\xi)-P_B(\eta,\xi)\parallel_2 \right ] \\
    &=\min_{\theta} \mathbb{E}_{\eta,\xi} \left [\parallel \mathcal{M}_{\theta}(P_{A}(\eta,\xi))-P_B(\eta,\xi)\parallel_2 \right ],
\label{eq:Loss_function}
\end{split}
\end{equation}
which is calculated by using the L2 norm, to ensure that the translated PDF $P_{AB}$ is indistinguishable from the target PDF $P_{B}$. The trained model is expected to be applied to cases, where the distance between $P_{A}$ and $P_{B}$ in PDF space is not much larger than the training case. Given that the input $P_{A}$ is based on the presumed joint PDF which is generally consistent with the target sub-grid distribution $P_{B}$, the scope of application is supposed to cover most combustion cases, thus the model's generalisability is enhanced. For this study, the joint PDF is calculated using the $\beta$-PDF distributions in Eq. (\ref{eq:joint_PDF}).

In summary, JPResUnet is trained using pairs of joint PDFs derived from the $\beta$-distribution and DNS as input and target, respectively, following the optimisation framework given in Eq. (\ref{eq:Loss_function}). A detailed discussion of data extraction and model structure follows in the subsequent sections.

\section{Data preprocessing and numerical setup} \label{sec:data_process}
\subsection{DNS cases} \label{subsec:dns_cases}
The JPResUnet is trained by using the DNS dataset of MILD combustion with varying mixture fractions and internal recirculation of exhaust gases (EGR). This combustion features a partially premixed combustion mode, indicated by the flame index observed across a wide field of the computational domain \cite{Doan2018,Minamoto2014}. In addition, a broad reaction zone is noted in MILD combustion \cite{Doan2018,Minamoto2014}, which implies significant subgrid-scale fluctuations in the thermo-chemical and -physical properties of the reacting mixture. These fluctuations are advantageous for training purposes, as they contribute to the development of a robust and generalised model.

The study focuses on two MILD combustion cases with different levels of dilution, labelled `AZ1' and `BZ1'. The initial thermo-chemical and -physical conditions for these cases are detailed in Table \ref{tab:MILD_details}. For both cases, the initial root-mean-square (RMS) value of the velocity fluctuation $u'_{\text{rms}}$ is around \qty{16.66}{m/s} and the integral length scale of the turbulence $\Lambda_0$ is around \qty{1.42}{mm}. Based on these quantities, the turbulent Reynolds number $\text{Re}_T$ and the Taylor micro-scale Reynolds number $\text{Re}_{\lambda}$ are 96 and 34.73, respectively. The ratio between the integral length scales of the progress variable and mixture fraction fields is $l_c/l_Z \approx 0.77$. The averaged progress variable is $\langle c \rangle=0.56$, and the RMS value of the initial fluctuation in the progress variable field is $\sigma_{c}/\langle c \rangle = 0.46$. Compared with the case AZ1, the dilution is enhanced for the case BZ1, where the oxygen level is reduced to $2\%$ by volume, inducing the difference in the mixture fraction field. The stoichiometric mixture fraction $Z_{\text{st}}$ for the case AZ1 and BZ1 are $0.01$ and $0.0058$, respectively. With a similar equivalence ratio, the averaged mixture fractions are $\langle Z \rangle = 0.008$ for the case AZ1 and $\langle Z \rangle = 0.0046$ for BZ1. The RMS values of the initial fluctuation of the mixture fraction $\sigma_{Z}/\langle Z \rangle$ for AZ1 and BZ1 are $1.05$ and $1.23$, respectively.  

The combustion behaviour varies significantly with different dilution levels. In AZ1, both thin and thick heat release zones, with a large variation in typical thickness, are observed, where the thickened zone is induced by the interaction of reaction zones. In contrast, BZ1 exhibits a much thicker heat release zone nearly over the whole computational domain, indicating more frequent interactions of reaction zones \cite{Doan2018}. To capture a wide range of combustion phenomena, training data is collected from the AZ1 case, while the BZ1 case is used for testing, as outlined in Table \ref{tab:case_information}.

\begin{tiny}
\begin{table}[h]
	\centering
	\caption{Initial conditions of the MILD combustion.}
    \resizebox{\linewidth}{!}{
	\begin{tabular}{c c c c c c c c c}
		\hline\noalign{\smallskip}
		Case & $\Lambda_0/l_Z$ & $l_c/l_Z$ & $X_{\text{O}_2}^{\text{max}}$ & $Z_{\text{st}}$ & $\langle Z \rangle$ & $\sigma_Z/\langle Z \rangle$ & $ \langle c \rangle$ & $\sigma_c/\langle c \rangle$ \\
		\noalign{\smallskip}\hline\noalign{\smallskip}
		AZ1 & 0.60 & 0.77 & 0.035 & 0.01 & 0.008 & 1.05 & 0.56 & 0.46\\
		\noalign{\smallskip}\hline\noalign{\smallskip}
        BZ1 & 0.60 & 0.77 & 0.020 & 0.0058 & 0.0046 & 1.23 & 0.56 & 0.46\\
		\noalign{\smallskip}\hline
	\end{tabular}}
	\label{tab:MILD_details}
\end{table}
\end{tiny}

A cubic computational domain of dimensions $L_x \times L_y \times L_z = 10 \times 10 \times$ \qty{10}{mm^3} was used, with 512 grid points in each spatial direction \cite{Doan2018}. The resulting grid size is $\delta x \approx$ \qty{20}{\mu m}, providing approximately 30 grid points within the smallest chemical thickness of methane--air combustion. Combustion chemistry is modelled using a modified chemical mechanism, MS-58, which is based on the Smooke and Giovangigli scheme \cite{Smoke1991} and has been enhanced to include $\textrm{OH}^*$ chemistry \cite{Kathrotia2012}. The MILD combustion was simulated using a DNS code, SENGA \cite{Cant2012}, with a timestep of $\delta t = \qty{1}{ns}$. After the first flow-through time, $\tau_f = L_x / U_{\text{in}}$, where $U_{\text{in}} = \qty{20}{m/s}$ representing the inflow bulk mean velocity, the initial transient exited the computational domain, and the simulation continued for another half of $\tau_f$ for data collection (approximately 60 snapshots). Further details regarding the numerical schemes, chemical mechanism, boundary conditions, and initial conditions can be found in \cite{Doan2018}.

\subsection{PDF extraction} \label{subsec:Data_extraction}
The PDFs are extracted from the sub-filter space within the DNS fields. For this study, the input PDFs are computed using a $\beta$-PDF in Eq. (\ref{eq:beta_PDF}), for given Favre filtered quantities, $\widetilde{c}$, $\widetilde{\sigma}_{c,\text{sgs}}^2$, $\widetilde{Z}$ and $\widetilde{\sigma}_{Z,\text{sgs}}^2$. The progress variable is defined using temperature, $c=(T-T_u)/(T_b(Z)-T_u)$, where $T_u=\qty{1500}{K}$ is the initial temperature for the unburnt mixture. The burnt mixture temperature, $T_b$, is calculated by using the local mixture fraction \cite{Minamoto2014}. Since combustion is adiabatic, this progress variable is equal to the one defined using species (with unity Lewis number) mass fractions, described previously and used for \textit{a posteriori} assessment. Hereinafter, the temperature-based progress variable is represented by $c_T$ to avoid ambiguity. 

The progress variable-related quantities are calculated as
\begin{equation}
  \widetilde{c_T}(\textbf{x},t) = \frac{1}{\overline{\rho}(\textbf{x},t)}\int_{\textbf{x}-\frac{\Delta}{2}}^{\textbf{x}+\frac{\Delta}{2}} \rho(\textbf{x}',t) c_T(\textbf{x}',t) \mathcal{G}(\textbf{x}') \,d\textbf{x}' \, ,
\label{eq:filtered_c}
\end{equation}
\begin{equation}
\begin{split}
  &\widetilde{\sigma}_{c_T,\text{sgs}}^2(\textbf{x},t) = \\
  &\frac{1}{\overline{\rho}(\textbf{x},t)}\int_{\textbf{x}-\frac{\Delta}{2}}^{\textbf{x}+\frac{\Delta}{2}} \rho(\textbf{x}',t) \left[c_T(\textbf{x}',t)-\widetilde{c_T}(\textbf{x},t) \right]^2 \mathcal{G}(\textbf{x}') \,d\textbf{x}' \, ,
\label{eq:variance_cT}
\end{split}
\end{equation}
where the prime $\textbf{x}'$ represents local position inside the filter of size $\Delta$, and $\mathcal{G}$ is the filter kernel. The Favre filtered mixture fraction and its variance are calculated using similar procedures on the DNS data. Two types of filter kernels are used: a spatial box filter and a Gaussian filter, and they are expressed as
\begin{align}
 &\mathcal{G}_{\text{box}}(\textbf{x}') = 
 \begin{cases}
 \frac{1}{\Delta} & \text{if}\; |\textbf{x}'-\textbf{x}| \leq \frac{\Delta}{2} \\
 0 & \text{otherwise}
 \end{cases} \\
&\mathcal{G}_{\text{Gaussian}}(\textbf{x}') = \Bigl(\frac{6}{\pi\Delta^2}\Bigr)^{\frac{1}{2}}\text{exp}\Big(-\frac{6\textbf{x}'^2}{\Delta^2}\Big).
 \label{eq:filter_kernel}
\end{align}
As noted in Table \ref{tab:case_information}, the normalised box filter kernel width of $\Delta^+=\Delta/\delta_{\text{th}}^{\text{st}}=1$ is employed for training. The term $\delta_{\text{th}}^{\text{st}}$ is the reference thermal thickness of stoichiometric laminar flame with the size of \qty{1.6}{mm} ($80\delta x$) and \qty{3}{mm} ($148\delta x$) for AZ1 and BZ1, respectively. Different filter sizes and Gaussian kernels are used during the testing phase.

The target joint distribution, denoted as $\widetilde{P}(\eta,\varphi;\textbf{x},t)$, is calculated by using the kernel density estimation (KDE) as
\begin{equation}
\widetilde{P}(\eta,\varphi;\textbf{x},t)=\frac{1}{nh}\sum_{i=1}^{n}K\left[\frac{(\eta,\varphi)-\frac{1}{\overline{\rho}(\textbf{x},t)}(\rho c_T,\rho \text{ln}\frac{Z}{Z_{\text{st}}})_i}{h} \right],
\label{eq:KDE}
\end{equation}
where $K$ and $h$ denote the kernel function and bandwidth, respectively. In this study, a bandwidth of 0.1 is used, and the Epanechnikov kernel \cite{Epanechnikov1969} is chosen for its computational efficiency, as it requires fewer samples $n$ only. This kernel is defined as
\begin{equation}
    K(s)=
    \begin{cases}
        \frac{3}{4}(1-s^2)\quad & \text{if}\; |s|\leq 1 \\
        0 & \text{otherwise}.
    \end{cases}
\label{eq:epanechnikov_kernel}
\end{equation}

The sample space variables $\eta$ and $\varphi$ in Eq. (\ref{eq:KDE}) are for the density-weighted progress variable and scaled mixture fraction, respectively. The scaled mixture fraction is expressed as $\widehat{Z}=\text{ln}(\widetilde{Z}/Z_{\text{st}})$ \cite{Yang2023}, which generates a relatively uniform distribution for flames with different flammability limits and stoichiometric values of the mixture fraction. The random variable space is discretised into a grid with dimensions $N_{\eta}\times N_{\varphi}=80 \times 100$. The progress variable dimension is linearly discretised, while the scaled mixture fraction dimension is split into two segments, $[-1.5, 0.3]$ and $[0.3, 1.8]$, with 65 and 33 points allocated respectively following earlier study \cite{Yang2023}. The prediction is transferred back to the $\eta$--$\xi$ space as $\widetilde{P}(\eta,\xi;\textbf{x},t)=\widetilde{P}(\eta,\varphi;\textbf{x},t)/{\xi}$ \cite{Yang2023}. For notational simplicity and clarity, $c_T$ and $Z$ are used directly instead of the sample-space variables for the PDFs, e.g., $\widetilde{P}(c_T,Z)$, $\widetilde{P}(c_T)$, $\widetilde{P}(Z)$, hereafter in this work.

It is noted that the joint distribution $\widetilde{P}(c_T,Z)$ obtained above is the filter density function (FDF) \cite{Pitsch2006}. The FDF is constructed from sub-filter space samples at a specific spatial location in a single DNS snapshot. Since DNS realisations are inherently unsteady, the FDF includes random variations \cite{Gao1993}, differing from the expected sub-grid PDF $\widetilde{P}(c_T,Z)$. To minimise this randomness, more samples from the sub-filter space would be required across multiple realisations with identical resolved fields \cite{Tong2001}, which is computationally expensive. Alternatively, this randomness can be significantly reduced if the training dataset for machine learning incorporates FDF samples collected over many realisations during a statistically stationary state \cite{Chen2021}. For this study, 25 DNS snapshots of the AZ1 MILD combustion case are selected. In each snapshot, the sub-filter space is systematically marched with a fixed spatial step to extract pairs of input-target PDFs. This process yields a training dataset of 33,275 samples, with $20\%$ reserved for validation to prevent overfitting.

\subsection{Machine learning algorithm} \label{subsec:model_description}

\begin{figure}[!ht]
     \centering
     \begin{subfigure}{0.5\textwidth}
         \centering
         \includegraphics[width=\textwidth]{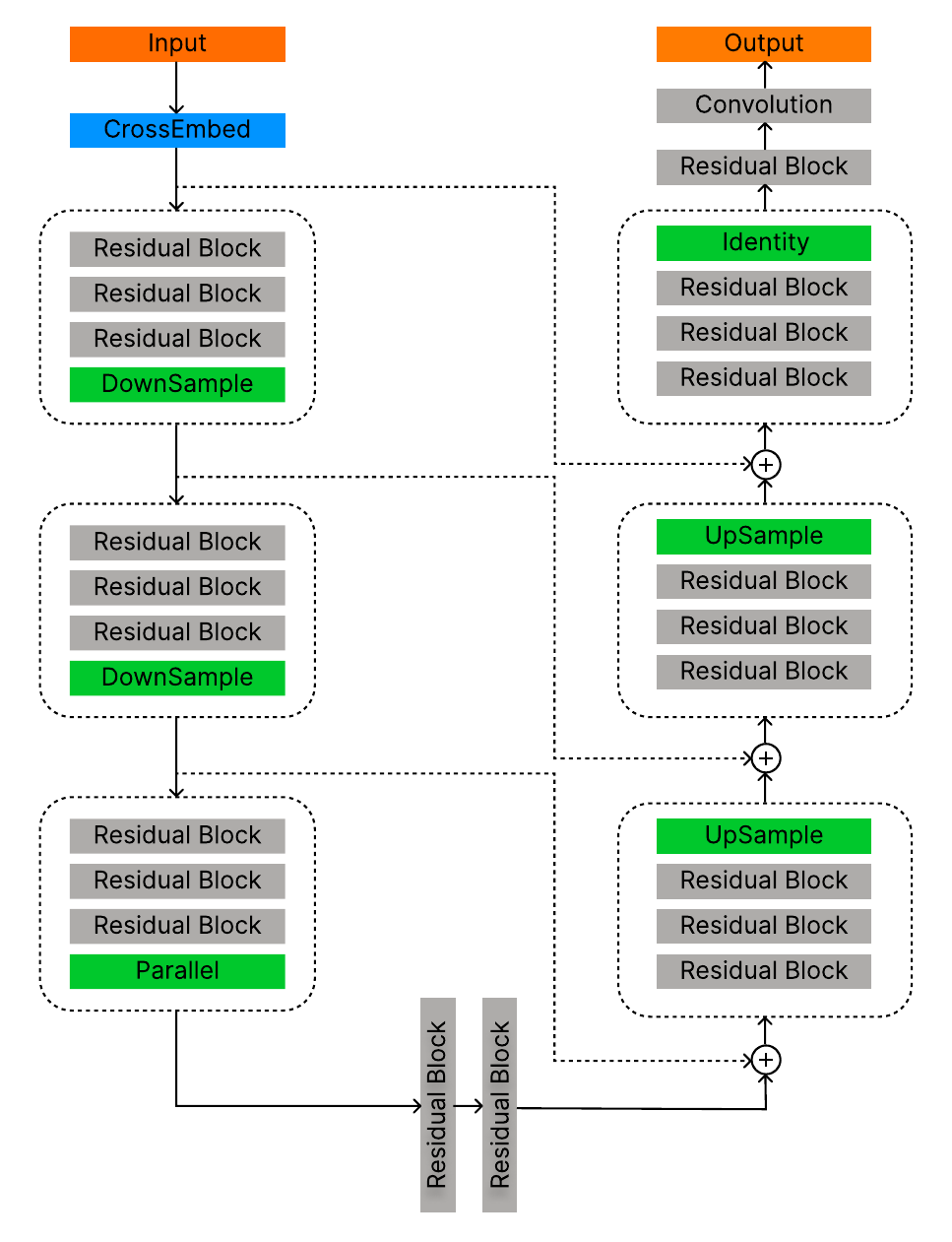}
         \caption{}
     \end{subfigure}
     \hfill
     \begin{subfigure}{0.4\textwidth}
         \centering
         \includegraphics[width=\textwidth]{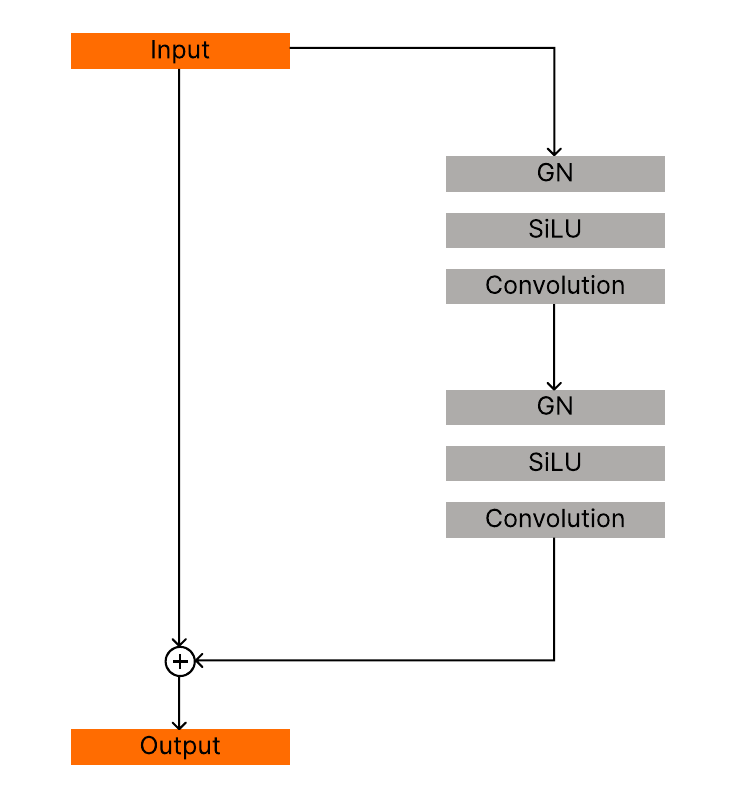}
         \caption{}
     \end{subfigure}
    \caption{\footnotesize Structure of (a) JPResUnet and (b) residual block.}
    \label{fig:Model_structure}
\end{figure}

The JPResUnet model features an encoder-decoder structure with skip connections, and each level within this structure utilises residual blocks, as shown in Fig. \ref{fig:Model_structure}a. Its architecture can be summarised as follows:
\begin{itemize}
\setlength\itemsep{0em}
    \item A cross-embedding layer and a residual block with an additional convolutional layer are deployed at the encoder-decoder's inlet and outlet, respectively. 
    \item The encoder and decoder are organised into three levels, each comprising three consecutive residual blocks followed by a size-changing unit. 
    \item Levels are connected via skip connections that integrate the input at each encoder level into the corresponding decoder level.
    \item Two residual blocks are positioned at the bottleneck of the model.
\end{itemize}

The cross-embedding layer (highlighted in blue in Fig. \ref{fig:Model_structure}a) adjusts the input to match the number of channels required by the encoder. This is achieved through three groups of convolutional kernels with sizes of $3 \times 3$, $7 \times 7$, and $15 \times 15$. The output from each kernel is padded to maintain the same size, and these outputs are concatenated along the channel dimension to reach the desired number of channels $N_{\text{ch}}$. The kernel allocation is divided into $N_{\text{ch}}/2$, $N_{\text{ch}}/4$ and $N_{\text{ch}}/4$ for the respective groups.

The residual blocks address the degradation issues often encountered in deep models by adding the input upstream to the residual function's output \cite{He2015}. This function includes a combination of group normalisation \cite{Wu2018}, a Sigmoid-Weighted Linear Unit (SiLU) \cite{Elfwing2018}, and a $3\times3$ convolution, which is repeated once, as illustrated in Fig. \ref{fig:Model_structure}b. The SiLU activation function, which outperforms the commonly used rectified linear unit (ReLU) in deep models by preventing inactive neurons during training, is defined as
\begin{equation}
  \text{SiLU}(x) = x * S(x),
  \label{eq:SiLU}
\end{equation}
where $S(x) = 1/(1+e^x)$ is the sigmoid function.

The size-changing unit, highlighted in green in Fig. \ref{fig:Model_structure}a, adjusts the input size to the required value. For an input with the shape $(N_{\text{ch}}^i,N_{c_T},N_{Z})$, the down-sampling unit changes it to $(N_{\text{ch}}^{i+1},N_{c_T}/2,N_{Z}/2)$ by using a $4\times4$ convolutional layer, where the superscript $i$ represents the model's level. The up-sampling unit, combining interpolation with a $3\times3$ convolution, restores the output to $(N_{\text{ch}}^{i+1},2N_{c_T},2N_{Z})$. A parallel unit uses two groups of convolutions with kernel sizes of $3\times3$ and $1\times1$ respectively, both producing outputs of shape $(N_{\text{ch}}^{i+1},N_{c_T},N_{Z})$, which are then summed.  

To preserve spatial information and facilitate gradient flow, skip connections are employed between the corresponding levels of the encoder and decoder. They scale the input at the $i^{\rm{th}}$ level of the encoder by $2^{-1/2}$ and concatenate it with the input at the corresponding decoder level along the channel dimension, resulting in a new input shape of $(N_{\text{ch}}^{i,\text{ec}}+N_{\text{ch}}^{i,\text{dc}},N_{c_T},N_{Z})$, where `ec' and `dc' denote the encoder and decoder respectively.

The JPResUnet's input PDF is a single-channel image sized at $80\times100$. The cross-embedding layer expands this into a shape of $(32,80,100)$. The encoder reduces the image resolution to $(128,20,25)$, and the decoder, which mirrors the encoder's structure, restores the resolution to $(32,80,100)$. The final residual block and convolution ensure the output PDF has the same size as the input. The detailed input-output information for every component of the JPResUnet is listed in \ref{appendix:structure}. It is noted that many iterations of the model's architecture have been tested, including the number of channels of the image generated by the cross-embedding layer, the number of levels in the encoder and decoder, and the number of residual blocks at each level. The current layout of the model shows the best performance and is adopted by the present study. 

The JPResUnet is implemented using PyTorch \cite{paszke2019}. Optimisation in Eq. (\ref{eq:Loss_function}) is handled by the AdamW algorithm \cite{loshchilov2019}, with a weight decay of $0.01$, an initial learning rate of $10^{-4}$, and a linear decay to $10^{-7}$ in 200 epochs. The batch size is set to 32. These hyperparameters were fine-tuned using the grid-search method. Training was halted at around 121 epochs as no further drop in validation loss was observed, taking about 6 hours on an NVIDIA GeForce RTX 4090 GPU.

\section{Performance of JPResUnet on different testing cases} \label{sec:priori}

The generality and effectiveness of the JPResUnet model were evaluated through a comparative study across different cases with varying levels of extrapolation, as outlined in Table \ref{tab:case_information}. To demonstrate the performance of JPResUnet, it was compared with an artificial neural network (ANN) consisting of three fully connected layers, with 256 and 512 neurons in the two hidden layers, respectively. This ANN was based on a previous work \cite{Chen2021} that successfully predicted the sub-grid PDF for MILD combustion. The ANN was modified for extrapolation using a methodology described in \cite{Yang2023}. The input features for the ANN included the Favre-filtered quantities $\widetilde{c}$, $\widetilde{g}_c$, $\widehat{Z}$ and $\widetilde{g}_Z$, while the output was the joint PDF $\widetilde{P}(c_T,\widehat{Z})$. Both the JPResUnet and ANN models were trained on the same dataset as described in Section \ref{subsec:Data_extraction} of the study.

\begin{table}[!ht]
    \centering
    \caption{List of training and testing information. The filter kernels "B" and "G" denote the box and Gaussian kernels, respectively}
    \begin{tabular}{l c c}
    \hline
    \rule{0pt}{15pt}\textbf{Data case} & \textbf{Filter kernels ($\Delta^+$)} & \textbf{Size of dataset}\\
    \hline
    \multicolumn{3}{c}{\rule{0pt}{15pt} \textit{Training}} \\[1ex]
        AZ1 &  B(1)  & 33275\\
    \hline
    \multicolumn{3}{c}{\rule{0pt}{15pt} \textit{In-sample validation}} \\[1ex]
        AZ1 &  B(1)  & 3993\\
    \hline
    \multicolumn{3}{c}{\rule{0pt}{15pt} \textit{Out-of-sample prediction: same combustion case}} \\[1ex]
        AZ1 &  B(1) B(1.5) B(2)  & 3993 3000 2187\\
        AZ1 & G(1) G(1.5) G(2) & 3993 3000 2187\\
    \hline
    \multicolumn{3}{c}{\rule{0pt}{15pt} \textit{Out-of-sample prediction: different combustion case}} \\[1ex]
        BZ1 & B(1) B(1.5) & 3000 1536\\
        BZ1 & G(1) G(1.5) & 3000 1536\\
    \hline
    \end{tabular}
    \label{tab:case_information}
\end{table}

\subsection{Validation: in-sample prediction}
The extrapolation capability of JPResUnet was tested on data that is slightly different from the training dataset. Within the temporal domain spanned by snapshots of the AZ1 flame used for training, three additional realisations of the flame were selected for in-sample predictions. 

\begin{figure}[H]
  \centering
  \includegraphics[width=\textwidth]{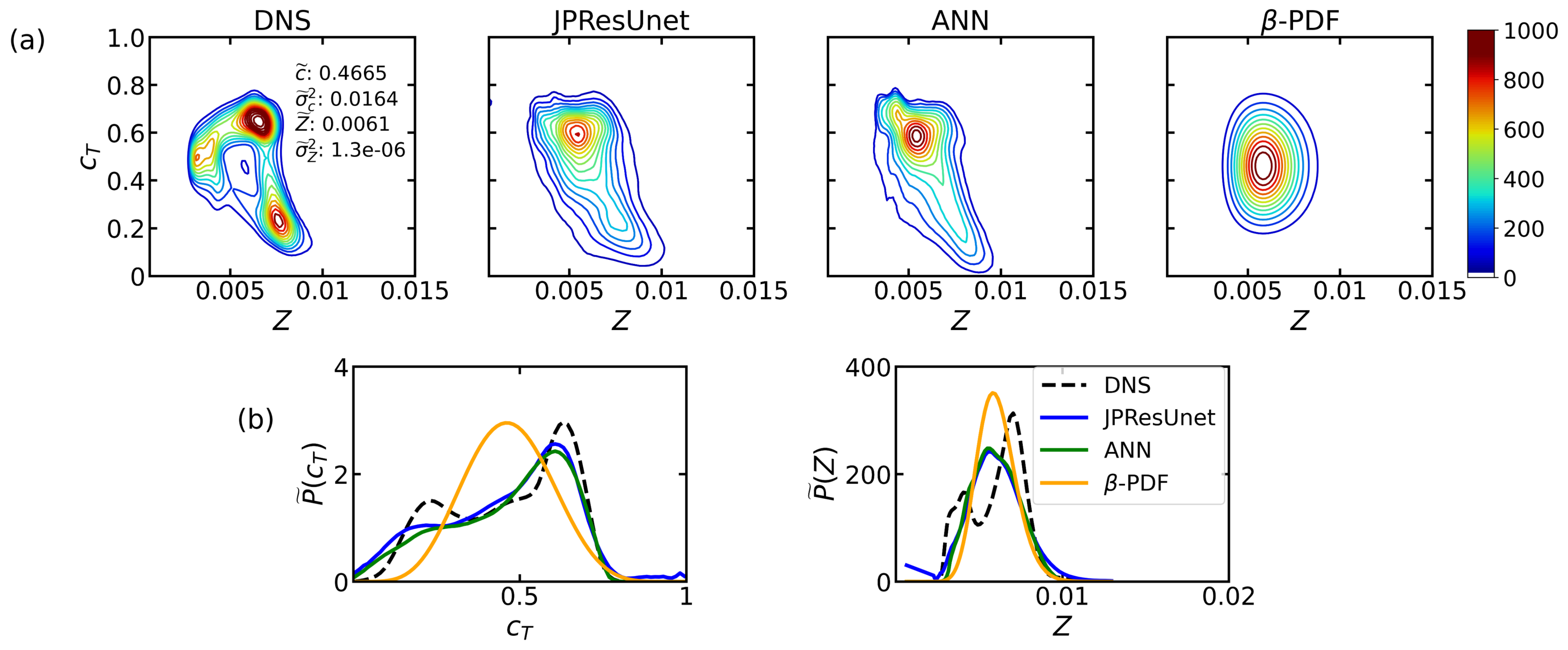}
  \caption{\footnotesize Comparative analysis of the joint and marginal PDFs between DNS and model in-sample predictions.}
  \label{fig:in-sample_PDFs}
\end{figure}

The JPResUnet model's ability to predict the joint PDF for a sub-grid space within the reaction region of AZ1 was compared with that of the ANN and the analytical model $\beta$-PDF. The results, presented in Fig. \ref{fig:in-sample_PDFs} of the study, include the marginal PDFs $\widetilde{P}(c_T)=\int\widetilde{P}(c_T,Z)\text{d}Z$ and $\widetilde{P}(Z)=\int\widetilde{P}(c_T,Z)\text{d}c_T$. JPResUnet captures the bi-modal distribution of the progress variable and a negative correlation between $c_T$ and $Z$, characteristic of MILD combustion under the AZ1 conditions \cite{Chen2021}, with contours closely matching DNS data. The ANN's prediction shows a similar agreement with the DNS, which is expected as many studies have verified such excellent in-sample prediction \cite{Chen2021, Yang2023}. In contrast, the analytic model, $\beta$-PDF, fails to capture the DNS data's shape and peak.

\begin{figure}[H]
	\centering
	\includegraphics[width=\textwidth]{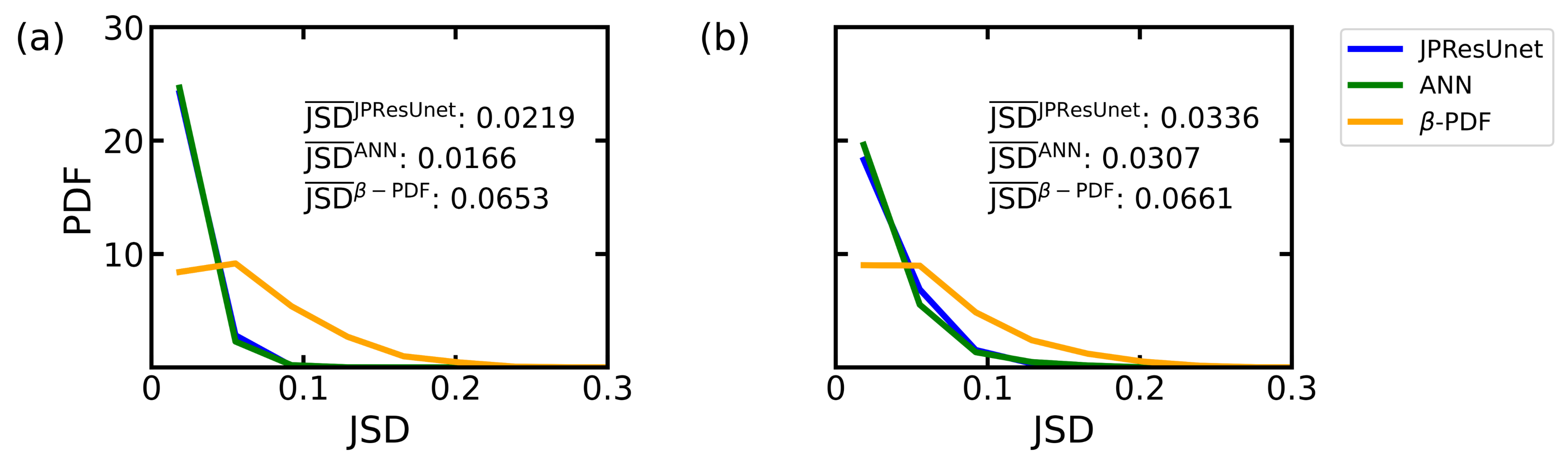}
	\caption{\footnotesize PDF of JSD for marginal PDFs of (a) progress variable $c_T$ and (b) mixture fraction $Z$, predicted on in-sample dataset.}
	\label{fig:in-sample_JSD}
\end{figure}

The overall accuracy of the JPResUnet model across the entire testing dataset (as listed in Table \ref{tab:case_information}) was assessed by comparing the predicted PDFs to those obtained from DNS using the Jensen-Shannon divergence (JSD) \cite{Endres2003}. The JSD measures the similarity between two distributions $P_1$ and $P_2$  and is calculated as 
\begin{equation}
  \textrm{JSD}(P_1||P_2) = \frac{1}{2}\sum_{n=1}^{N} \left (P_1(n)\ln \frac{P_1(n)}{P_2(n)} + P_2(n)\ln \frac{P_2(n)}{P_1(n)} \right ),
  \label{eq:JSD_function}
\end{equation}
where $N$ denotes the total discretised points in the random variable space.  The JSD value is bounded between $0$ and $\text{ln}(2)$, with smaller values indicating higher similarity between the two distributions. The PDFs from the model and DNS were taken as $P_1$ and $P_2$, respectively, and the JSD values for the marginal PDFs of the progress variable and the mixture fraction were calculated and plotted in Fig. \ref{fig:in-sample_JSD}.

Both the JPResUnet and ANN models exhibited high accuracy compared to the $\beta$-PDF model, as their JSD plots for $\widetilde{P}(c_T)$ and $\widetilde{P}(Z)$ are clustered closer to zero. The mean JSD values for these models were approximately one-third and one-half of the corresponding values for the $\beta$-PDF model, respectively. This finding reaffirms the superiority of the ANN model for in-sample predictions. The JPResUnet's similar performance to the ANN confirms its accuracy, though further improvements in in-sample prediction are beyond the scope of this study.

\subsection{Testing with different filter widths and kernels}

Due to constraints in computational and experimental resources, the availability of high-fidelity data for turbulent flames is currently limited, which makes it challenging to comprehensively cover practical scenarios, especially complex geometries. This limitation necessitates development of models that can be applied to scenarios significantly different from the training dataset. To validate this capability, the JPResUnet model was evaluated using data sampled from realisations of AZ1 at temporal steps beyond the range used for training. Additionally, since most LES utilise an implicit filter, where the filter type and size are unknown, the model was tested on data derived using various filter kernels and sizes to ensure its practical applicability.

\begin{figure}[!htbp]
	\centering
	\includegraphics[width=\textwidth]{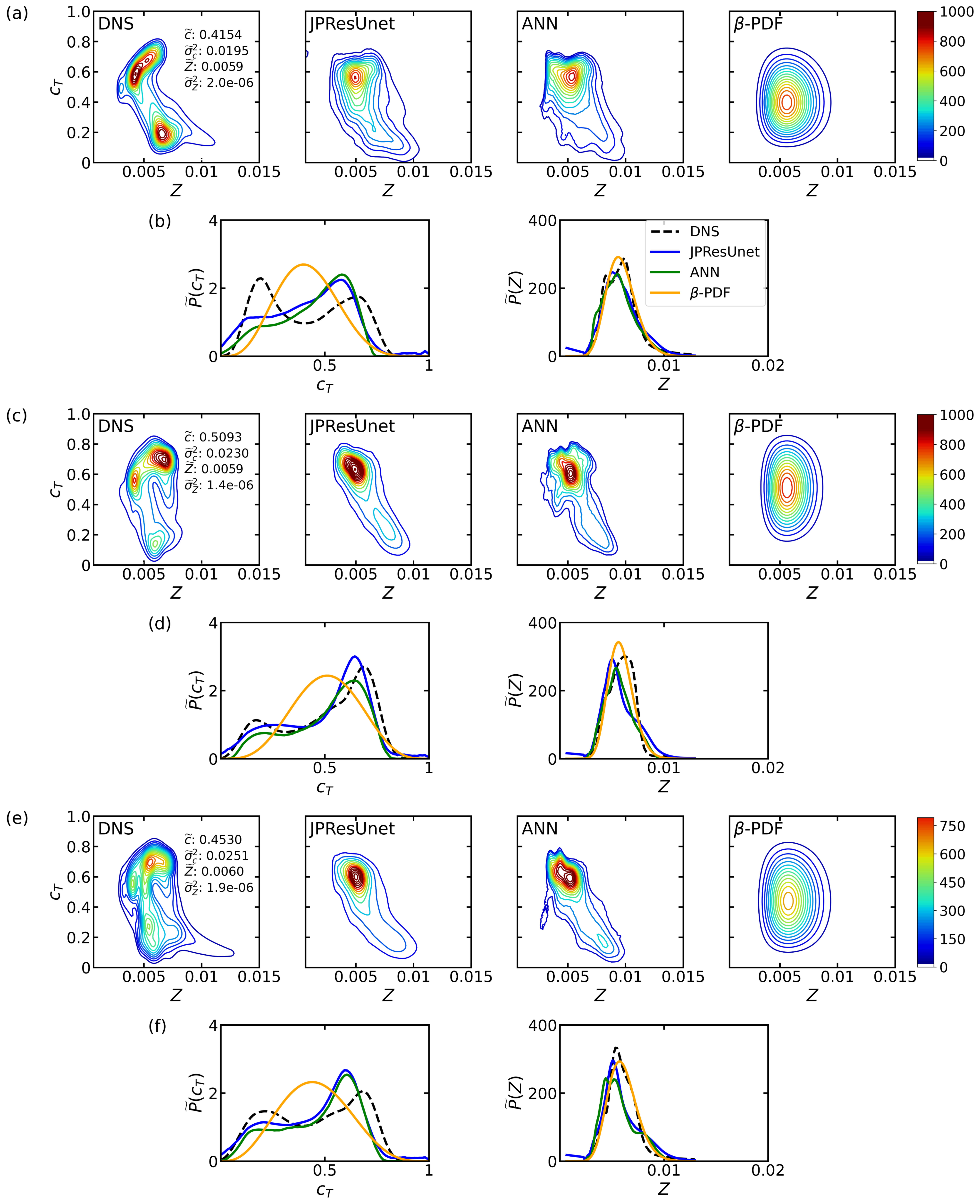}
	\caption{\footnotesize Comparative analysis of the joint and marginal PDFs between DNS and model out-of-sample predictions for the case AZ1, utilising a box filter with widths of (a)-(b) $\Delta^+=1$, (c)-(d) $\Delta^+=1.5$ and (e)-(f) $\Delta^+=2$.}
	\label{fig:out-sample_AZ1_box_PDFs}
\end{figure}

\begin{figure}[!htbp]
	\centering
	\includegraphics[width=\textwidth]{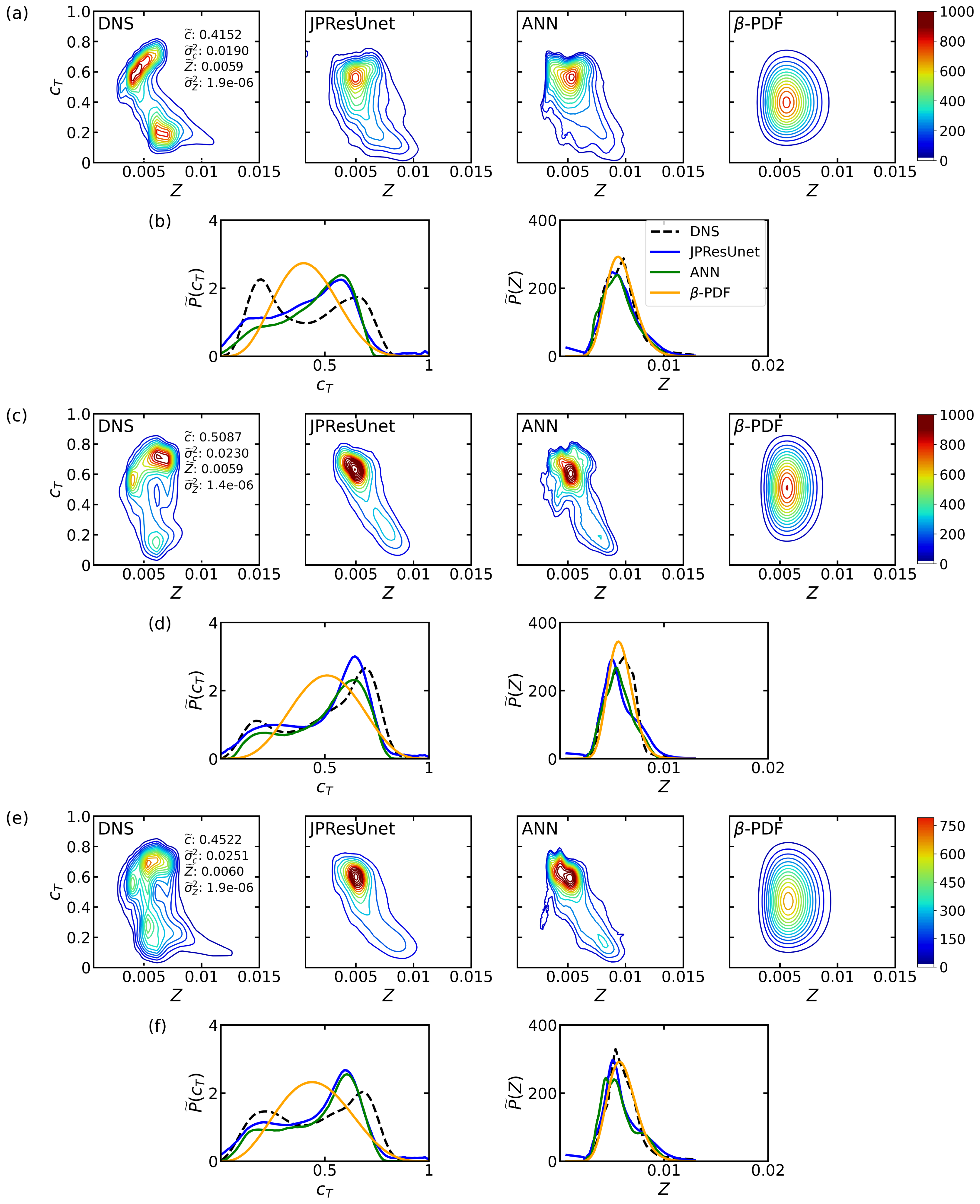}
	\caption{\footnotesize Comparative analysis of the joint and marginal PDFs between DNS and model out-of-sample predictions for the case AZ1, utilising a Gaussian filter with widths of (a)-(b) $\Delta^+=1$, (c)-(d) $\Delta^+=1.5$ and (e)-(f) $\Delta^+=2$.}
	\label{fig:out-sample_AZ1_gaussian_PDFs}
\end{figure}

The performance of the models was evaluated using data extracted with box and Gaussian filter kernels of widths $\Delta^+=1$, $1.5$, and $2$. Larger filter widths were not considered due to the size of DNS domain, as $\Delta^+=3$ approaches half the side length of the domain, resulting in an insufficient number of samples extracted from selected snapshots. Moreover, large filters fail to encompass a sufficient range of the turbulent kinetic energy, rendering the validated models impractical for \textit{a posteriori} simulations and testing, as discussed in \cite{Chen2021}. The predicted joint and marginal PDFs for a local sub-grid space with the box and Gaussian kernels at the three filter sizes are presented in Fig. \ref{fig:out-sample_AZ1_box_PDFs} and \ref{fig:out-sample_AZ1_gaussian_PDFs}, respectively. 

The predicted, both joint and marginal, PDFs are observed to be insensitive to the choice of the filter kernel, specifically for $\Delta^+=1$. At this width, there is minimal variation in filtered quantities, as shown in Fig. \ref{fig:out-sample_AZ1_box_PDFs}a and \ref{fig:out-sample_AZ1_gaussian_PDFs}a. Consequently, the models, including the ANN and $\beta$-PDF approach, generate consistent predictions despite the subtle input variations. The JPResUnet's predictions also remain constant since the inputs from the $\beta$-distribution under both box and Gaussian filters are the same. Therefore, the discussion on local sub-grid PDF predictions focuses on the results obtained using the box filter only (Fig. \ref{fig:out-sample_AZ1_box_PDFs}).

The JPResUnet demonstrates excellent performance, accurately reproducing the shape and peak of joint PDF contours across all filter widths, with close agreement with DNS results. This accuracy is particularly evident in $\widetilde{P}(c_T)$ at larger filter widths (as shown in Fig. \ref{fig:out-sample_AZ1_box_PDFs}d and \ref{fig:out-sample_AZ1_box_PDFs}f). In contrast, the $\beta$-PDF approach fails to provide satisfactory predictions except for the marginal distribution of the mixture fraction. This analytical model generates statistical distributions, as the Gaussian-like shape observed in the $\widetilde{P}(c_T)$ plots, while missing the instantaneous features. This deficiency compromises the accuracy of the PDF-based approach in modelling instantaneous reaction rates, as will be further demonstrated later in this section.

The ANN model performs comparably to JPResUnet but shows noticeable under-predictions in marginal PDFs as filter width increases (as seen in Fig. \ref{fig:out-sample_AZ1_box_PDFs}f). Additionally, the joint PDF contours predicted by ANN exhibit wrinkled shapes, attributed to the unsteadiness of the instantaneous sub-grid PDFs (FDFs) used during training. JPResUnet avoids these issues by leveraging the statistical structure of $\beta$-PDF while preserving sub-grid complexity, resulting in smooth and consistent predictions across filter sizes. This robustness underscores JPResUnet’s high generalisation capability.

\begin{figure}[!htbp]
	\centering
	\includegraphics[width=\textwidth]{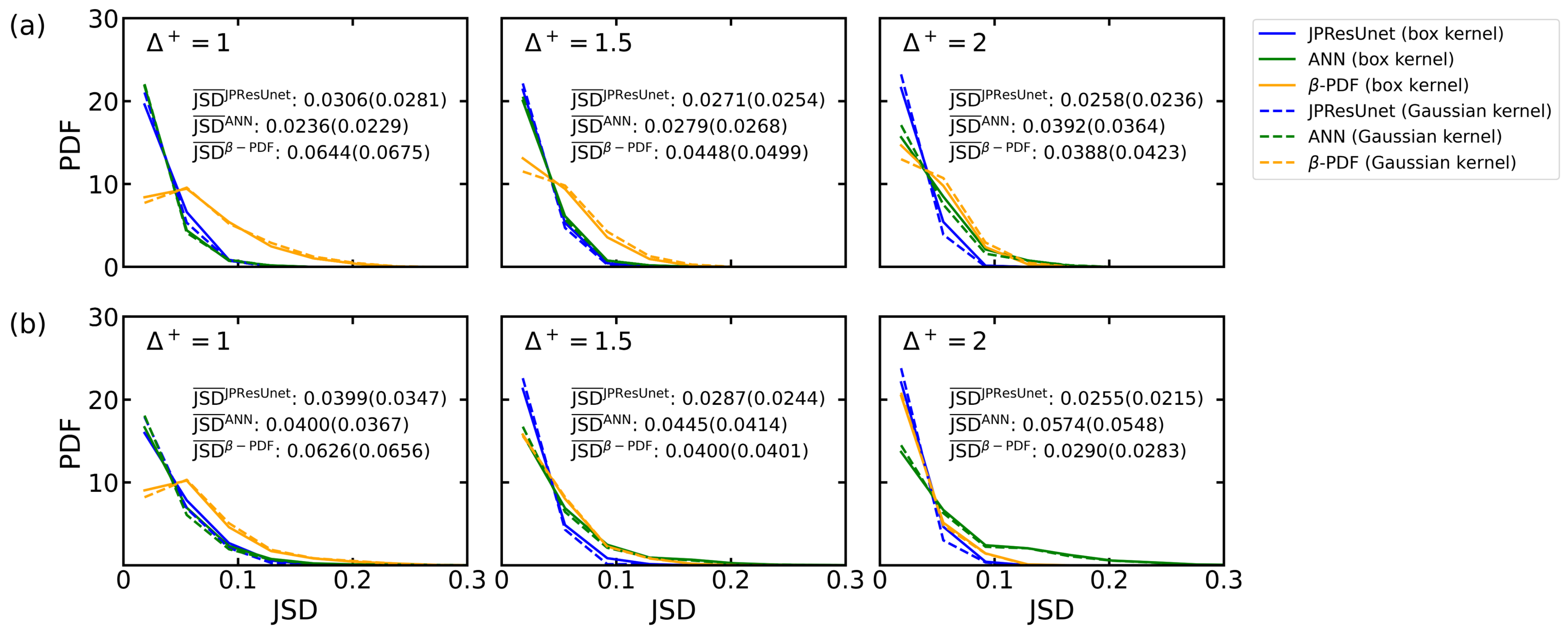}
	\caption{\footnotesize PDF of JSD for marginal PDF of (a) progress variable $c_T$ and (b) mixture fraction $Z$ for $\Delta^+=1$, $1.5$ and $2$, predicted on out-of-sample dataset of the case AZ1. Solid and dashed lines correspond to the box and Gaussian filters, respectively. The mean JSD values for models' prediction based on the box filter (Gaussian filter) are also listed.}
	\label{fig:out-sample_AZ1_box_JSD}
\end{figure}

Figure \ref{fig:out-sample_AZ1_box_JSD} presents the PDF of the JSD values for model predictions at varying filter widths and kernels (solid and dashed lines correspond to the box and Gaussian filters, respectively). The mean JSD values for each model's predictions with the box (and also Gaussian) filter are listed. JPResUnet demonstrates consistent high accuracy across all filter widths and types, with over $80\%$ of its JSD values concentrated below $0.05$ and decreasing mean JSD value as filter width increases. This reduction indicates the benefit of translating the statistical $\beta$-distribution, which shows a decline as well. While ANN performs well at $\Delta^+=1$, its accuracy decreases with filter width, leading to the highest mean JSD values at $\Delta^+=2$, reflecting the negative influence of randomness in the training data.

\begin{figure}[!htbp]
	\centering
	\includegraphics[width=\textwidth]{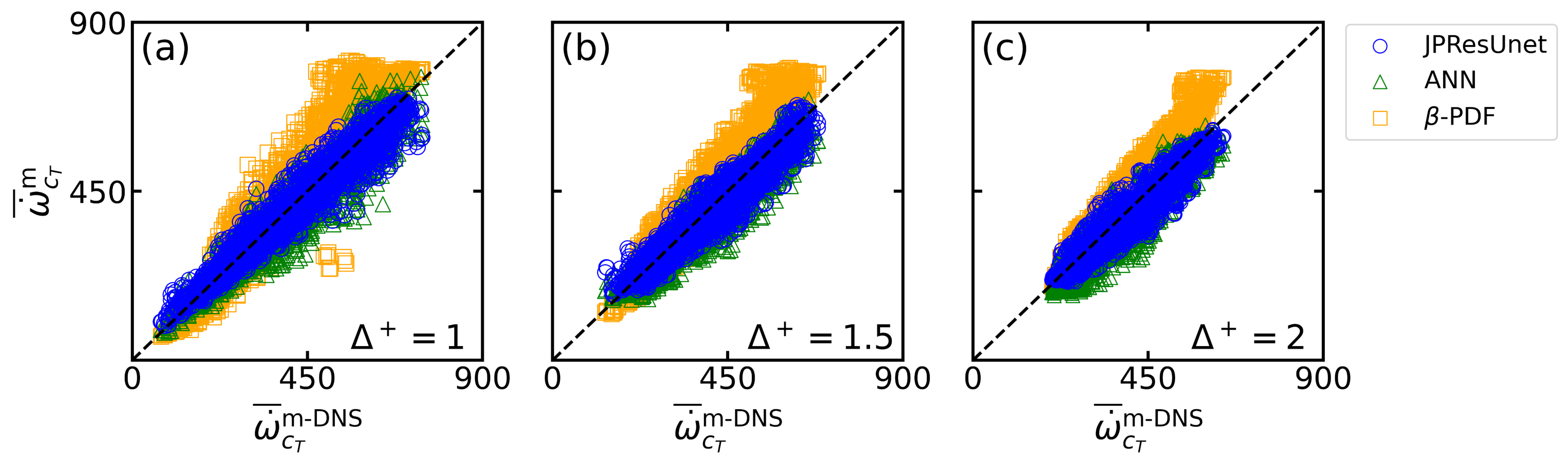}
	\caption{\footnotesize Scatter plot of the progress variable source term modelled by joint PDFs from different models and from the DNS for the case AZ1, using the box filter with different filter widths.}
	\label{fig:out-sample_AZ1_box_reaction_rate}
\end{figure}

Using the predicted joint PDF, the filtered reaction rate source term $\overline{\dot{\omega}}_{c_T}$ is calculated similarly to Eq. (\ref{eq:premixed_reaction_rate}), where the flamelet reaction rate and density in the integrand are replaced with the doubly conditional averaging counterparts $<\dot{\omega}_{c_T}/\rho|c_T,Z>$ over the DNS data (60 snapshots in total), as verified in \cite{Chen2021}. The reaction rate of $c_T$ is calculated as $\dot{\omega}_{c_T}=\dot{q}/[c_p(T_b-T_u)]$, where $\dot{q}$ and $c_p$ are the volumetric heat release rate and mixture specific heat capacity respectively. The comparison between the filtered reaction rates calculated by using the joint PDF from models $\overline{\dot{\omega}}_{c_T}^{\text{m}}$ and from DNS $\overline{\dot{\omega}}_{c_T}^{\text{m-DNS}}$ with different filter widths is illustrated in the scatter plot shown in Fig. \ref{fig:out-sample_AZ1_box_reaction_rate}.  Due to the similarity of the joint PDFs predicted using the box and Gaussian filters, the modelled filtered reaction rates are very similar, so only the results for the box filter are shown here (the results for the Gaussian filter are in the supplementary material).

Among all the models, JPResUnet demonstrates the best performance across all filter sizes, accurately modelling the filtered reaction rate closely clustered around the diagonal without noticeable over- or under-prediction. Similarly, ANN provides comparable predictions with substantial overlap with JPResUnet. However, as filter size increases, ANN's predictions become more scattered, reflecting a decline in predictive performance discussed previously. The analytical $\beta$-PDF approach produces reasonable estimates but consistently over-predicts the reaction rate $\overline{\dot{\omega}}_{c_T}^\text{m}$ across all filter widths, aligning with prior observations \cite{Chen2021}.

The accuracy of filtered reaction rate modelling is assessed using the root-mean-square of a normalised error (RMSE), defined as
\begin{equation}
    \text{RMSE}^{\text{m}} = \sqrt{\frac{1}{N} \sum_{i=1}^{N}\left(\frac{\overline{\dot{\omega}}_{c_T,i}^\text{m}-\overline{\dot{\omega}}_{c_T,i}^\text{m-DNS}}{\overline{\dot{\omega}}_{c_T,i}^\text{m-DNS}} \right)^2},
    \label{eq:RMSE}
\end{equation}
where the index $i$ denotes each sample in the test dataset with the size $N$. The RMSE values for each model with the box (and Gaussian) filter at all filter widths are listed in Table \ref{tab:RMSE_AZ1}. The RMSE values for JPResUnet are the lowest for all filter widths and remain below 0.1 for both box and Gaussian kernels. By contrast, the RMSE value for the ANN increases significantly with the filter width, reaching around 1.5 times the value of JPResUnet's predictions. Although the $\beta$-PDF approach shows improved performance at higher filter widths, its RMSE remains the highest, due to the over-prediction observed in Fig. \ref{fig:out-sample_AZ1_box_reaction_rate}c. 

\begin{table}[!ht]
    \centering
    \caption{Root-mean-square error (RMSE) for filtered reaction rate by using different models for AZ1 with the box (Gaussian) filter at different filter widths.}
    \resizebox{0.8\linewidth}{!}{
      \begin{tabular}{c c c c}
      \hline
      \textbf{Model} & \textbf{$\Delta^+=1$} & \textbf{$\Delta^+=1.5$} & \textbf{$\Delta^+=2$}\\
      \hline
      JPResUnet & 0.0868 (0.0849) &  0.0904 (0.0864)  & 0.0982 (0.0950)\\
      \hline
      ANN &  0.101 (0.0999)  & 0.112 (0.107) & 0.135 (0.130)\\
      \hline
      $\beta-$PDF &  0.165 (0.169) & 0.152 (0.157) & 0.147 (0.150)\\
      \hline
      \end{tabular}}
    \label{tab:RMSE_AZ1}
\end{table}

\subsection{Extrapolating to a higher dilution level}

The extrapolation capability of the JPResUnet model has been rigorously assessed and shown to be robust across different filter widths, demonstrating improved accuracy compared to both the state-of-the-art ML model and the widely used analytical model. To further evaluate this capability, JPResUnet was tested on a distinct MILD combustion case, BZ1, which exhibits different thermo-chemical and -physical characteristics from the training case AZ1 (detailed in Section \ref{subsec:dns_cases}). The model evaluations were conducted using data extracted with two filter widths, $\Delta^+=1$ and $1.5$.

The local sub-grid distributions within the reaction zone obtained using the box kernel are depicted in Fig. \ref{fig:out-sample_BZ1_box_PDFs}. The results for the Gaussian kernel are similar, which are included in \ref{appendix:Gaussian_filter}. Unlike the AZ1 case, the DNS joint PDFs for BZ1 display a less pronounced bimodal distribution and a more dispersed shape, indicating a more distributed reaction zone. JPResUnet accurately captures these characteristics across all filter widths, with its predictions aligning closely with DNS. In contrast, the ANN model predicts the joint PDF contour with a discontinuity at $\Delta^+=1$, as evidenced by fluctuations in the marginal PDF for the mixture fraction (Fig. \ref{fig:out-sample_BZ1_box_PDFs}b), and significantly underpredicts the marginal PDF for the progress variable. At the larger filter width, ANN fails to produce a reasonable PDF, with the output resembling a highly concentrated delta-distribution, as illustrated in Figs. \ref{fig:out-sample_BZ1_box_PDFs}c and \ref{fig:out-sample_BZ1_box_PDFs}d. These results underscore the excellent generality of JPResUnet. By leveraging the PDF-translation mechanism introduced in Section \ref{subsec:PDF_translation}, it is well-suited for cases with a reasonable statistical distribution of the sub-grid space. Conversely, ANN's performance is limited by its training dataset, resulting in poor predictions when the test case deviates significantly from the training data.

\begin{figure}[!htbp]
	\centering
	\includegraphics[width=\textwidth]{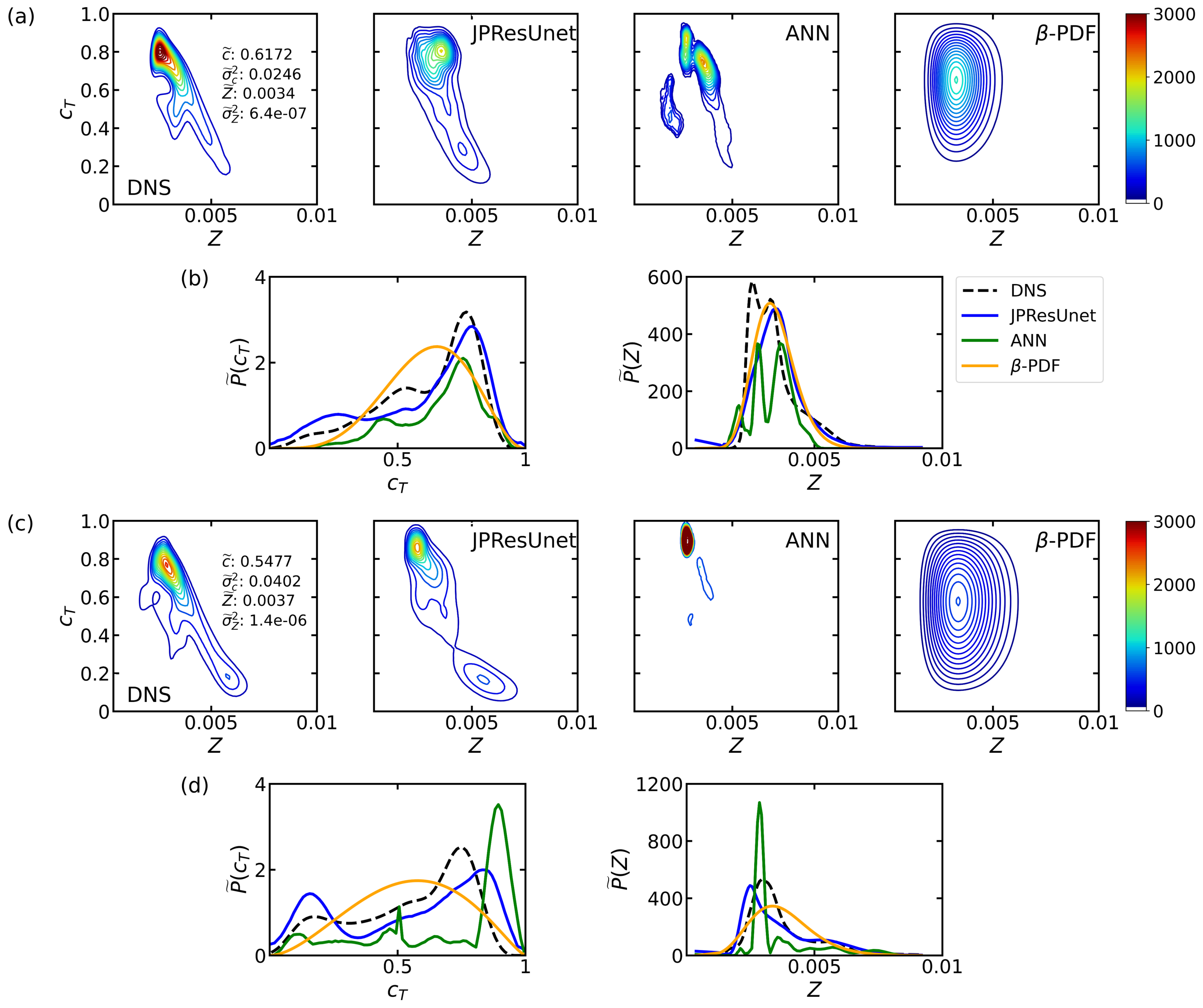}
	\caption{\footnotesize Comparative analysis of the joint and marginal PDFs between DNS and model predictions for the case BZ1, utilising a box filter with widths of (a)-(b) $\Delta^+=1$ and (c)-(d) $\Delta^+=1.5$.}
	\label{fig:out-sample_BZ1_box_PDFs}
\end{figure}

\begin{figure}[!htbp]
	\centering
	\includegraphics[width=\textwidth]{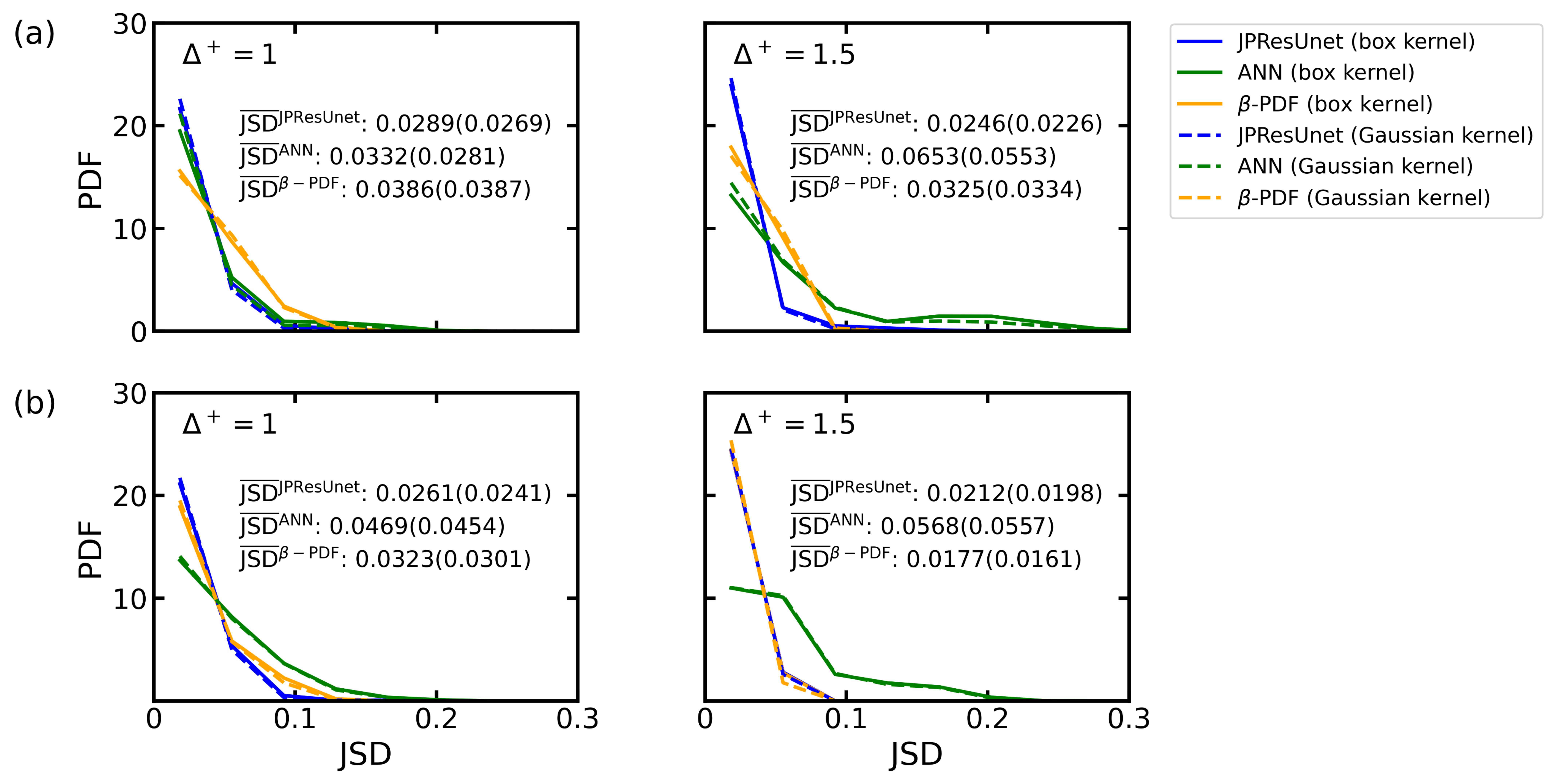}
	\caption{\footnotesize PDF of JSD for marginal PDF of (a) progress variable $c_T$ and (b) mixture fraction $Z$ for $\Delta^+=1$ and $1.5$, predicted on the case BZ1. Solid and dashed lines correspond to the box and Gaussian filters, respectively. The mean JSD values for models' prediction based on the box filter (Gaussian filter) are also listed.}
	\label{fig:out-sample_BZ1_two_filters_JSD}
\end{figure}

The overall accuracy of the models over the test dataset is compared using JSD of the PDF predictions in Fig. \ref{fig:out-sample_BZ1_two_filters_JSD}. Similar to the previous observation (Fig. \ref{fig:out-sample_AZ1_box_JSD}), JPResUnet achieves the highest predictive accuracy, with over $80\%$ JSD values below 0.05, leading to the lowest mean value (less than $0.03$). In contrast, the ANN model exhibits a significant number of JSD values exceeding 0.1, particularly at $\Delta^+=1.5$, where its mean JSD values are double those of JPResUnet, reflecting a low-fidelity prediction.

The filtered reaction rate modelled by different approaches with the box filter is compared to DNS results in Fig. \ref{fig:out-sample_BZ1_box_reaction_rate}, with results for the Gaussian filter provided in \ref{appendix:Gaussian_filter}. Among all models, JPResUnet delivers the most accurate predictions, which are tightly clustered around the diagonal, despite slight underprediction at $\Delta^+=1.5$. In contrast, the ANN model exhibits significant under-predictions due to inaccurate PDF predictions, while the $\beta$-PDF approach shows over-prediction for the region of intense reaction. The RMSE values presented in Table \ref{tab:RMSE_BZ1} confirm the very good performance of JPResUnet, remaining the lowest RMSE (around 0.1). By comparison, the ANN model’s inferior performance is quantitatively evidenced by high RMSE values, exceeding 0.2 at $\Delta^+=1.5$.

\begin{figure}[!htbp]
	\centering
	\includegraphics[width=0.8\textwidth]{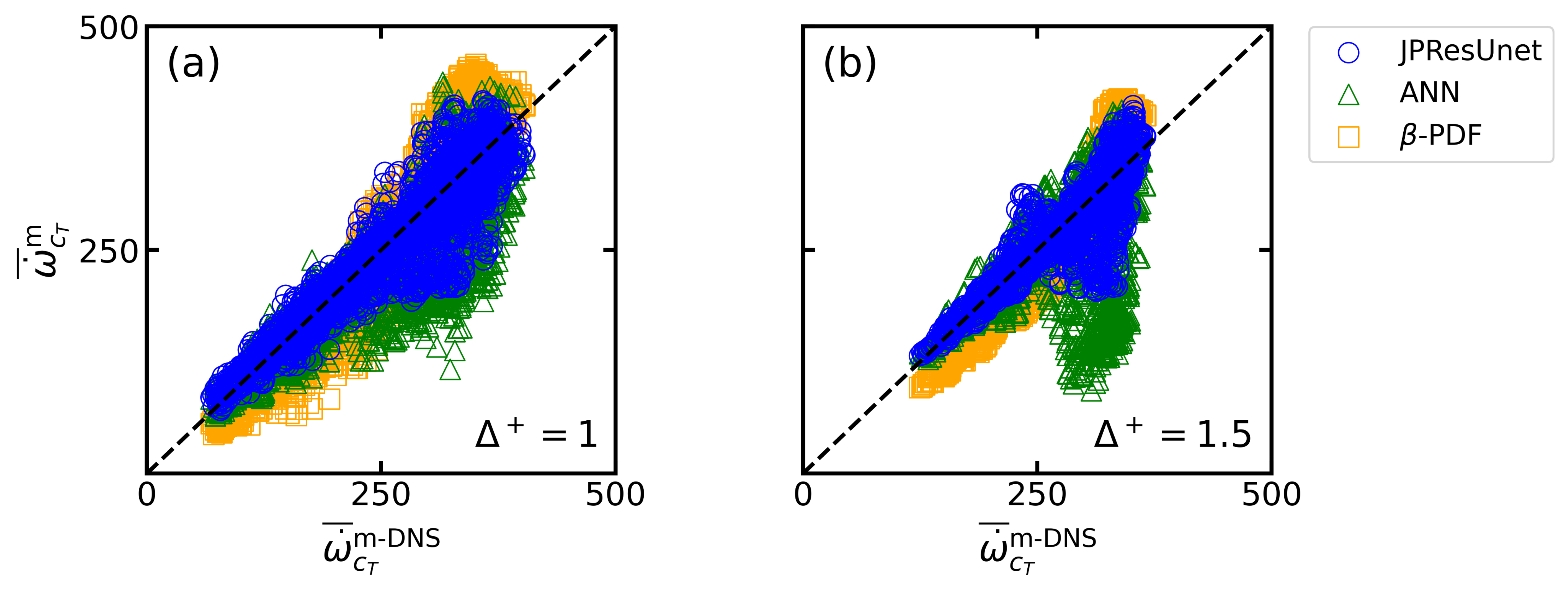}
	\caption{\footnotesize Scatter plot of the progress variable source term modelled by joint PDFs from different models and from the DNS for the case BZ1, using the box filter with different filter widths.}
	\label{fig:out-sample_BZ1_box_reaction_rate}
\end{figure}

\begin{table}[!ht]
    \centering
    \caption{Root-mean-square error (RMSE) for filtered reaction rate by using different models for BZ1 with the box (Gaussian) filter at different filter widths.}
    \resizebox{0.6\linewidth}{!}{
      \begin{tabular}{c c c}
      \hline
      \textbf{Model} & \textbf{$\Delta^+=1$} & \textbf{$\Delta^+=1.5$} \\
      \hline
      JPResUnet & 0.103 (0.102) &  0.0997 (0.0989)\\
      \hline
      ANN &  0.156 (0.151)  & 0.244 (0.238)\\
      \hline
      $\beta-$PDF &  0.175 (0.179) & 0.131 (0.134)\\
      \hline
      \end{tabular}}
    \label{tab:RMSE_BZ1}
\end{table}

\section{\textit{a posteriori} assessment} \label{sec:posteriori}
\begin{sloppypar}
The preceding section illustrates the performance of the JPResUnet model in predicting sub-grid PDFs for various MILD combustion scenarios within an \textit{a priori} assessment framework. This performance is benchmarked against the ANN and $\beta$-distribution approach on out-of-sample filter widths and kernels. Notably, JPResUnet maintains consistently high predictive accuracy across these conditions. Consequently, the model is evaluated in an \textit{a posteriori} LES to substantiate its robustness and potential in practice.
\end{sloppypar}

\subsection{LES setup} \label{subsec:les}
\begin{figure}[!htbp]
	\centering
	\includegraphics[width=0.8\textwidth]{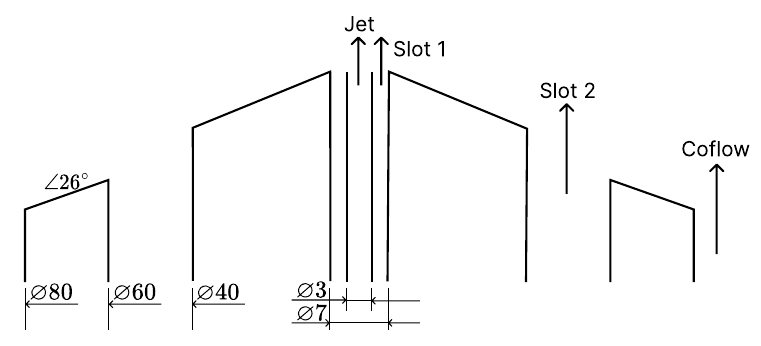}
	\caption{\footnotesize Schematic of the multi-regime burner (MRB) with diameters in mm.}
	\label{fig:MRB}
\end{figure}

The experimental setup of the multi-regime burner (MRB) is shown schematically in Fig. \ref{fig:MRB}. This involves three streams with different flow rates and equivalence ratios, resulting in inhomogeneous mixing of reactants downstream of the central jet. The bluff body positioned between slots 1 and 2 generates a recirculation zone composed of burnt products, stabilising the inner and outer flames. A high-velocity, fuel-rich premixed methane--air mixture is issued through the central jet, while pure air is supplied through slot 1. A range of equivalence ratios of the central premixed jet and bulk-mean air velocity for slot 1 is considered in the experimental study of \cite{Butz2019}. One of those cases, MRB26b, is considered for the \textit{a posteriori} testing of the ML models presented in the previous section. This specific case is considered because it showed clear multi-regime combustion including local mixtures having mixture fraction values beyond the flammability limits \cite{Massey2023b}. The equivalence ratio of the central methane--air jet is 2.6, with a velocity of \qty{105}{m/s}. The bulk-mean air velocity is \qty{15}{m/s} for slot 1. The lean premixed methane--air mixture with an equivalence ratio of 0.8 flows through slot 2 at a velocity of \qty{20}{m/s}. The flame is shielded from the external disturbances by using a low-velocity air co-flow around the burner. The temperatures of the mixtures introduced through the jet, slot 1, slot 2, and the co-flow are \qty{309}{}, \qty{333}{}, \qty{307}{}, and \qty{298}{K} respectively. Further details on the burner configuration and measurement techniques are discussed by Butz et al. \cite{Butz2019,Butz2022}.

The numerical setup follows the previous study \cite{Massey2023b}. The computational domain is cylindrical, with a radius of \qty{600}{mm} and a length of \qty{600}{mm} from the exit of the central jet nozzle, which corresponds to approximately five times the flame length. This domain is discretised using about 3.5M hexahedral numerical cells. The finest resolution is applied within the central jet region, which has the highest velocity. The mesh size ranges from 0.1 to \qty{0.3}{mm} in the inner flame region. The outer flame region has cell sizes ranging from 0.4 to \qty{1}{mm}. The velocity boundary conditions at the inlets utilise an inflow turbulence generator based on the synthetic eddy method \cite{Jarrin2006}. A three-dimensional steady RANS with the Reynolds stress equation model is conducted to obtain the mean profiles of velocity, the Reynolds stress tensor, and the streamwise integral length scale required for the LEMOS inflow generator \cite{Kornev2007}. Wave transmissive boundary conditions are imposed on the outflow boundaries to prevent acoustic wave reflection. All burner geometry walls are set to be no-slip and adiabatic.

In the \textit{a priori} assessment, JPResUnet has demonstrated its effectiveness in capturing the instantaneous features of sub-grid distributions, indicating the potential for real-time inference during the simulation. However, the primary challenge associated with on-the-fly deployment is the computational cost. The numerous nonlinear operations inherent in ML models significantly increase the computational burden compared to widely used tabulation methods, which rely on linear interpolation within a relatively low-dimensional look-up table (LUT). The computational costs increase with model complexity. Consequently, a tabulation approach was employed as a practical compromise with tables generated using ML model, JPResUnet, tested in the previous section. The filtered reaction rate source terms $\overline{\dot{\omega}}_\text{p}$ and $\overline{c\dot{\omega}^*}$ modelled by using JPResUnet were stored in a new LUT, which were retrieved during the simulation. Future research will explore more promising alternatives, such as integrating JPResUnet in LES using GPU acceleration for on-the-fly inference.

The simulations are performed using OpenFOAM v7 with a modified PIMPLE algorithm (rhoPimpleFoam solver). Second-order central difference schemes are used for velocity, and an implicit Euler scheme is employed for time marching, with a small variable time step on the order of $\mathcal{O}(10^{-7}\text{s})$ to ensure the CFL number remains below 0.4 across the entire domain. Time-averaged statistics are obtained over a period of 25 flow-through times, which is necessary to achieve convergence due to the presence of low-velocity, large-scale structures within the recirculation zone. The flow-through time is based on the size of the recirculation zone, the distance upstream to slot 2, and the velocity of slot 2; one flow-through time is approximately \qty{2}{ms}, and five flow-through times are required for the flames to stabilise after ignition. The simulation is run on ARCHER2, a UK high-performance computing facility, using 1920 cores for \qty{62}{h} of wall clock time.

\subsection{Results} \label{subsec:LES_results}
The results of \textit{a posteriori} assessment of the JPResUnet are compared with the experiment as well as the prior LES study \cite{Massey2023b} employing a $\beta$-PDF-based LUT (referred to as $\text{Look-up}_{\beta}$). It is worth noting that the case $\text{Look-up}_{\beta}$ uses the PDF with the resolution of $400\times500$ in the $c\times Z$ space for the reaction rate modelling in Eq. (\ref{eq:premixed_reaction_rate}), significantly finer than the resolution of $80\times100$ for JPResUnet. To investigate the effect of the resolution, a higher-resolution version with $384\times384$ is also used ($400\times500$ is not used because of excessive demand for computer memory). Hereinafter, results from this higher-resolution JPResUnet are denoted with the subscript `hr', while those from the lower-resolution JPResUnet tested in Section \ref{sec:priori} are denoted with the subscript `lr'. 

\begin{figure}[!htbp]
	\centering
	\includegraphics[width=\textwidth]{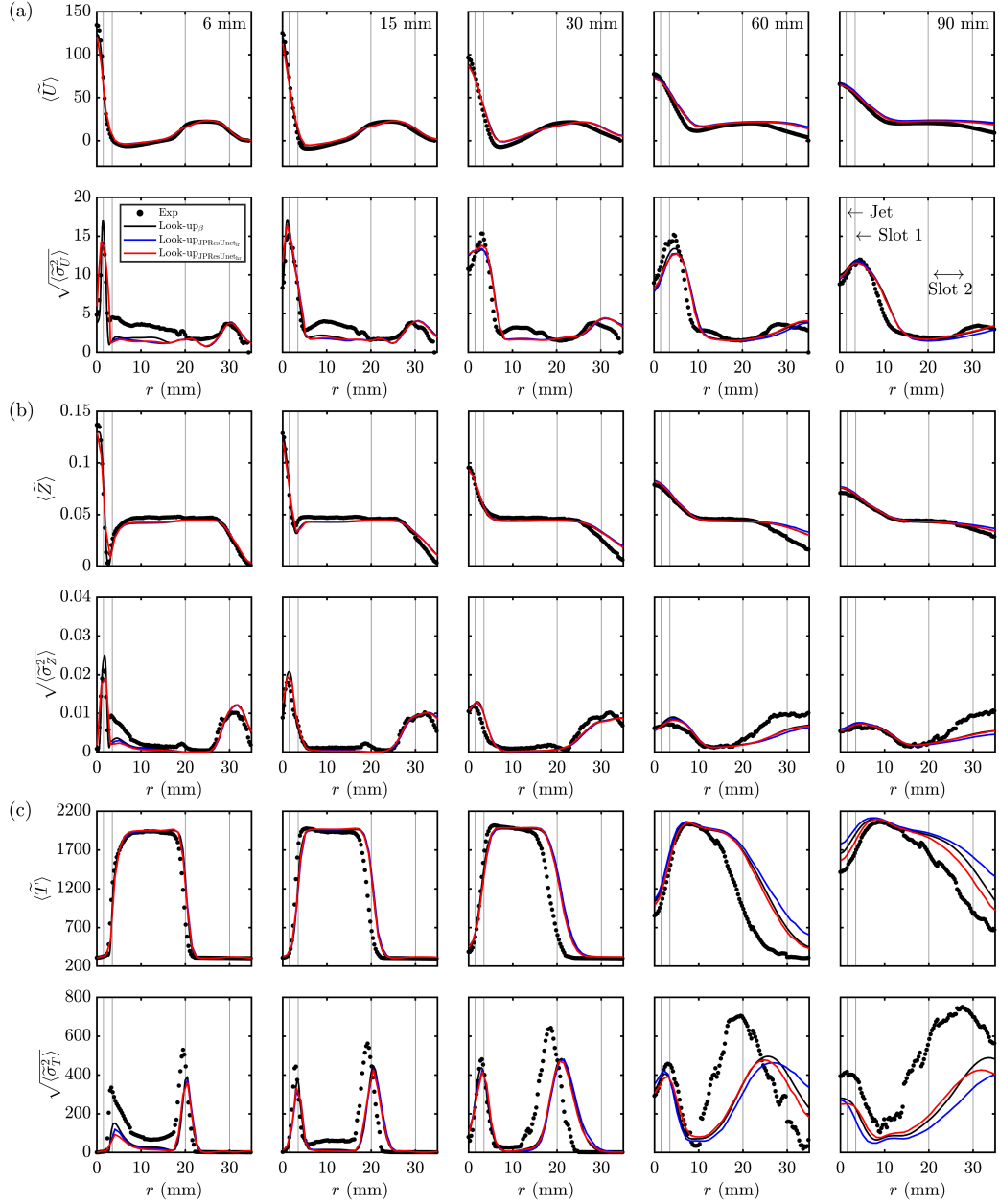}
	\caption{\footnotesize Radial profiles at different streamwise locations of the (a) axial velocity, (b) mixture fraction, and (c) temperature for the simulations (lines) and measurements (symbols).}
	\label{fig:MRB_vel_mix_tem}
\end{figure}

Figure \ref{fig:MRB_vel_mix_tem} presents the radial profiles of time-averaged and root-mean-squared (RMS) axial velocity, mixture fraction, and temperature at various streamwise locations, with all results averaged in the azimuthal direction. The velocity field is not affected by PDF models, with mean and RMS axial velocities (Fig. \ref{fig:MRB_vel_mix_tem}a) showing indistinguishable profiles across all approaches and aligning well with experimental data. The mixture fraction profiles (Fig. \ref{fig:MRB_vel_mix_tem}b) exhibit similar trends, with all approaches producing comparable results that closely match experimental data in the inner reaction region and over- and under-predict the mean and RMS values, respectively, for the outer shear layer of slot 2. Notably, the sensitivity to PDF resolution of JPResUnet is detected, with the lower-resolution JPResUnet (represented by blue lines) showing slightly more deviations than other simulations.

Regarding the temperature field (Fig. \ref{fig:MRB_vel_mix_tem}c), all simulations exhibit similar radial profiles that align well with experimental data in the near-field region (below \qty{30}{mm}). Further downstream (60 to \qty{90}{mm}), simulations show a good agreement with the measurement around the jet region with a slight over-prediction of the mean value. For the outer flame ($r\geq \qty{15}{mm}$), the mean temperature is over-predicted, indicating insufficient air entrainment \cite{Massey2023b}, and the peak of the RMS value is shifted towards a higher radial position. The high-resolution JPResUnet (represented by red lines) improves agreement with experimental data, particularly at \qty{90}{mm}, where it reduces both the mean temperature over-prediction and RMS peak shift. Conversely, the low-resolution model amplifies these deviations, underscoring the critical role of PDF resolution in achieving accurate predictions. 

Overall, with proper resolution in the PDF space, JPResUnet demonstrates very good accuracy in \textit{a posteriori} assessments, especially in regions where the $\beta$-PDF model struggles. It is stressed here again that the current deployment of the JPResUnet is a compromise due to the limited computational resources. As a model trained by an instantaneous dataset, on-the-fly inference should be conducted in future to further validate and strengthen the above findings.

\section{Conclusion} \label{sec:conclusion}
This study developed a novel PDF-translation model, JPResUnet, inspired by image-to-image translation techniques and utilising a residual U-net architecture. The model predicts the sub-grid joint probability density function (PDF) of the progress variable and mixture fraction in partially premixed flames, leveraging the analytical $\beta$-PDF as input. This approach ensures consistent translation within the PDF space during training and testing, enhancing the model's generalisability across diverse applications. Training was conducted using direct numerical simulation (DNS) data from the methane--air Moderate or Intense Low-oxygen Dilution (MILD) combustion, with a box filter at unit-normalised width. 

The performance of JPResUnet was first validated through an \textit{a priori} assessment against a well-studied Artificial Neural Network (ANN) and the $\beta$-PDF approach. On in-sample data, JPResUnet achieved predictive accuracy comparable to the ANN while outperforming the $\beta$-distribution, as demonstrated by the contours of local sub-grid distributions and Jensen-Shannon Divergence (JSD) values.

For out-of-sample data with varying filter kernels and widths, JPResUnet provided consistently accurate predictions of local sub-grid PDFs and produced smoother contours compared to ANN, reducing sensitivity to unsteadiness in the training data. It also demonstrated good performance in modelling the filtered reaction rate source term, achieving the lowest root-mean-square error (RMSE) across all filter widths, while the ANN performance declined with increasing filter widths.

\begin{sloppypar}
JPResUnet's generalisability was further assessed in the case of MILD combustion with a higher dilution level than the training case at two filter widths. JPResUnet's predictions closely matched the DNS results, while the ANN's predictions deteriorated significantly to entire failure at $\Delta^+=1.5$. The JSD values for JPResUnet remained the lowest, while those for ANN increased to approximately twice that of JPResUnet at the large filter width. JPResUnet also aligned closely with DNS in modelling the filtered reaction rate source term, with the lowest RMSEs, in contrast to ANN's substantial under-prediction, particularly at $\Delta^+=1.5$.
\end{sloppypar}

An \textit{a posteriori} assessment of JPResUnet was conducted through a large eddy simulation (LES) of the multi-regime burner (MRB). To address the computational cost of on-the-fly inference, JPResUnet was implemented via a look-up table (LUT), and compared against experimental data and a conventional $\beta$-PDF-based LUT. An additional JPResUnet with a higher PDF resolution ($384 \times 384$ in $c\times Z$) was tested to assess the influence of resolution. While the velocity and mixture fraction fields showed negligible differences, the high-resolution JPResUnet reduced deviations in the temperature field, particularly in the outer reaction region at downstream locations. In contrast, increased deviations were observed for low-resolution JPResUnet, implying the performance is sensitive to the PDF resolution.

In conclusion, JPResUnet demonstrates robust performance across various combustion scenarios, surpassing traditional methods in accurately capturing complex features and exhibiting better generality compared to the ANN. Its ability to reduce deviations in both \textit{a priori} and \textit{a posteriori} assessments underscores the potential for LES applications. Future research will focus on optimising computational efficiency to enhance the model’s applicability in practice.

\section*{Acknowledgements}
J.C.M. and N.S. acknowledge the financial support from Mitsubishi Heavy Industries, Ltd., Takasago, Japan. This work used the ARCHER2 UK National Supercomputing Service ( https://www.archer2.ac.uk ). The authors are grateful to ARCHER2 for the financial and computational allocation as a part of an ARCHER2 Pioneer Project (e808). 

\bibliographystyle{elsarticle-num-CNF} 
\bibliography{library} 

\appendix

\section{Detailed JPResUnet structure}
\label{appendix:structure}

\begin{table}[H]
	\centering
	\caption{\footnotesize Network structure of JPResUnet.} 

   {\fontsize{7}{8}\selectfont
    \resizebox{\linewidth}{!}{
	\begin{tabular}{c c c c c c c}
	\hline\hline\noalign{\smallskip}
		 & Unit level & Level components & \parbox[c]{2cm}{\centering Filter\\
         (size/number)} & Stride & Padding & \parbox[c]{3cm}{\centering Output size\\ ($\text{channels}\times\text{height}\times\text{width}$)} \\
		\noalign{\smallskip}\hline\noalign{\smallskip}
  
		Input & & & & & & $1\times80\times100$ \\
		\noalign{\smallskip}\hline\noalign{\smallskip}
  
        CrossEmbed & & & \parbox[c]{2cm}{\centering \(\begin{aligned}3&\times3/16\\ 7&\times7/8\\ 15&\times15/8\end{aligned}\)} & \parbox[c]{1cm}{\centering 1\\ 1\\ 1} & \parbox[c]{1cm}{\centering 1\\ 3\\ 7} & $32\times80\times100$\\
		\noalign{\smallskip}\hline\noalign{\smallskip}
  
        \multirow{13}{*}{Encoder} & \multirow{4}{*}{Level 1} & Residual Block & $3\times3/32$ & 1 & 1 & $32\times80\times100$ \\  &                          & Residual Block & $3\times3/32$ & 1 & 1 & $32\times80\times100$ \\  &                          & Residual Block & $3\times3/32$ & 1 & 1 & $32\times80\times100$ \\  &                          & DownSample & $4\times4/32$ & 2 & 1 & $32\times40\times50$ \\ \noalign{\smallskip}\cline{2-7}\noalign{\smallskip} & \multirow{4}{*}{Level 2} & Residual Block & $3\times3/32$ & 1 & 1 & $32\times40\times50$ \\       &                          & Residual Block & $3\times3/32$ & 1 & 1 & $32\times40\times50$ \\       &                          & Residual Block & $3\times3/32$ & 1 & 1 & $32\times40\times50$ \\       &                          & DownSample & $4\times4/64$ & 2 & 1 & $64\times20\times25$ \\ \noalign{\smallskip}\cline{2-7}\noalign{\smallskip} & \multirow{5}{*}{Level 3} & Residual Block & $3\times3/64$ & 1 & 1 & $64\times20\times25$ \\       &                          & Residual Block & $3\times3/64$ & 1 & 1 & $64\times20\times25$ \\       &                          & Residual Block & $3\times3/64$ & 1 & 1 & $64\times20\times25$ \\\noalign{\smallskip}       &                          & Parallel & \parbox[c]{2cm}{\centering $3\times3/128$\\ $1\times1/128$} & \parbox[c]{1cm}{\centering 1\\ 1} & \parbox[c]{1cm}{\centering 1\\ 0} & $128\times20\times25$ \\
        \noalign{\smallskip}\hline\noalign{\smallskip}

        \multirow{2}{*}{Bridge} & Level 4 & Residual Block & $3\times3/128$ & 1 & 1 & $128\times20\times25$ \\ \noalign{\smallskip}\cline{2-7}\noalign{\smallskip} & Level 5 & Residual Block & $3\times3/128$ & 1 & 1 & $128\times20\times25$ \\
        \noalign{\smallskip}\hline\noalign{\smallskip}
        
        \multirow{13}{*}{Decoder} & \multirow{4}{*}{Level 6} & Residual Block & $3\times3/128$ & 1 & 1 & $128\times20\times25$ \\  &                          & Residual Block & $3\times3/128$ & 1 & 1 & $128\times20\times25$ \\  &                          & Residual Block & $3\times3/128$ & 1 & 1 & $128\times20\times25$ \\  &                          & UpSample & $3\times3/64$ & 1 & 1 & $64\times40\times50$ \\ \noalign{\smallskip}\cline{2-7}\noalign{\smallskip} & \multirow{4}{*}{Level 7} & Residual Block & $3\times3/64$ & 1 & 1 & $64\times40\times50$ \\       &                          & Residual Block & $3\times3/64$ & 1 & 1 & $64\times40\times50$ \\       &                          & Residual Block & $3\times3/64$ & 1 & 1 & $64\times40\times50$ \\       &                          & UpSample & $3\times3/32$ & 1 & 1 & $32\times80\times100$ \\ \noalign{\smallskip}\cline{2-7}\noalign{\smallskip} & \multirow{4}{*}{Level 8} & Residual Block & $3\times3/32$ & 1 & 1 & $32\times80\times100$ \\       &                          & Residual Block & $3\times3/32$ & 1 & 1 & $32\times80\times100$ \\       &                          & Residual Block & $3\times3/32$ & 1 & 1 & $32\times80\times100$ \\       &                          & Identity &  &  & & $32\times80\times100$ \\
        \noalign{\smallskip}\hline\noalign{\smallskip}

         & & Residual Block & $3\times3/32$ & 1 & 1 & $32\times80\times100$ \\
        \noalign{\smallskip}\hline\noalign{\smallskip}
         & & Convolution & $3\times3/1$ & 1 & 1 & $1\times80\times100$ \\
        \noalign{\smallskip}\hline\noalign{\smallskip}
        Output & & & & & & $1\times80\times100$ \\
        \noalign{\smallskip}\hline\hline
	\end{tabular}}}
    \label{tab:JPResUnet_detailed_structure}
	
\end{table}

\section{Results for the Gaussian filter}
\label{appendix:Gaussian_filter}

\begin{figure}[H]
	\centering
	\includegraphics[width=0.8\textwidth]{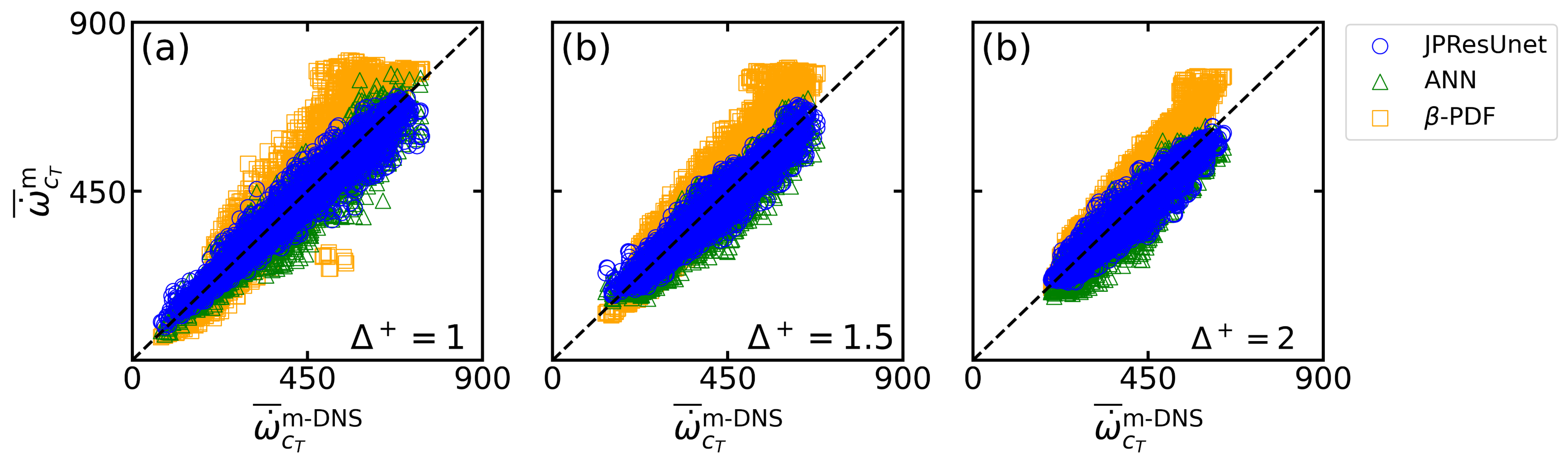}
	\caption{\footnotesize Scatter plot of the progress variable source term modelled by joint PDFs from different models and from the DNS for the case AZ1, using the Gaussian filter with different filter widths.}
	\label{fig:out-sample_AZ1_gaussian_reaction_rate}
\end{figure}

\begin{figure}[H]
	\centering
	\includegraphics[width=0.8\textwidth]{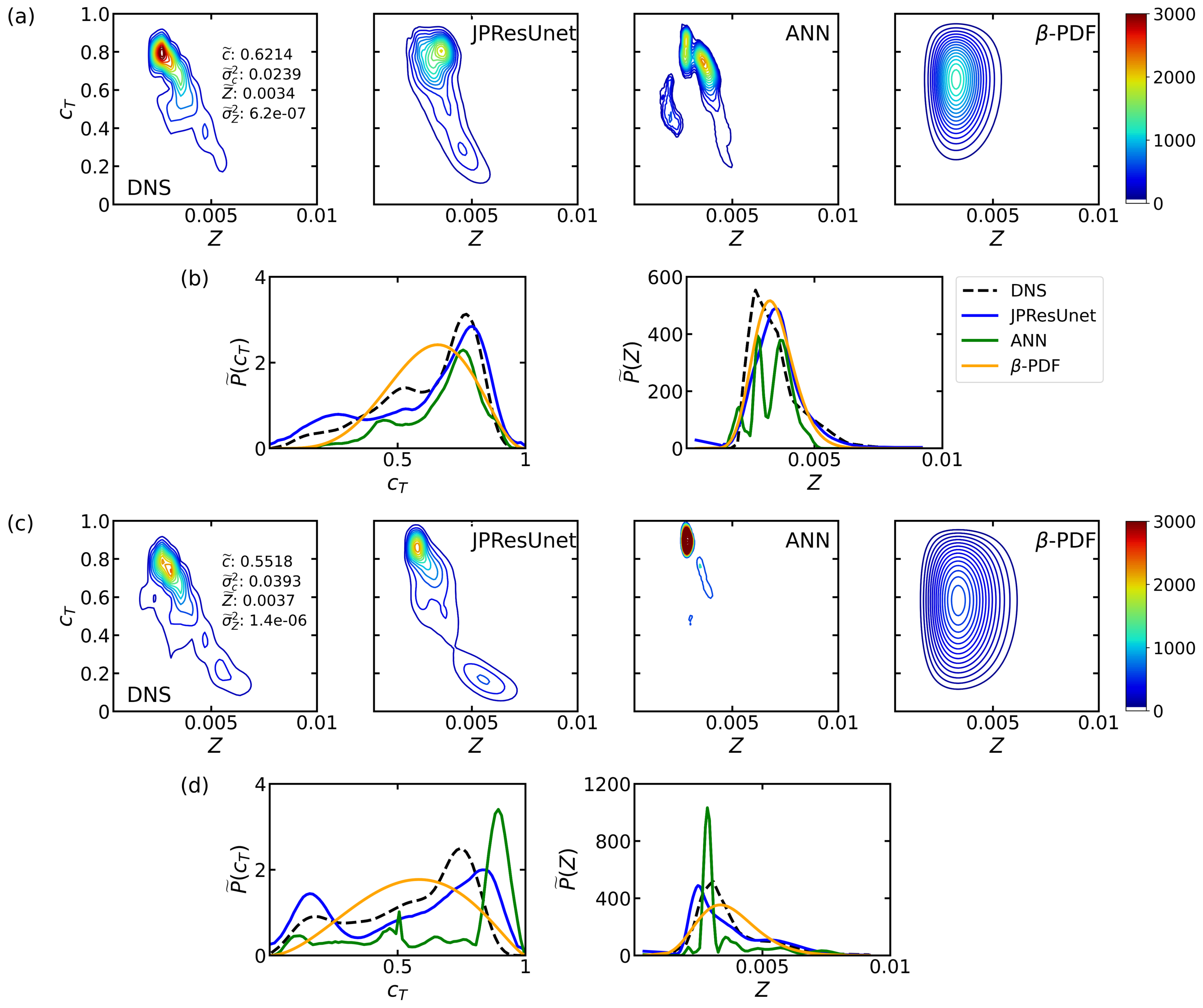}
	\caption{\footnotesize Comparative analysis of the joint and marginal PDFs between DNS and model predictions for the case BZ1, utilising a Gaussian filter with widths of (a)-(b) $\Delta^+=1$, (c)-(d) $\Delta^+=1.5$.}
	\label{fig:out-sample_BZ1_gaussian_PDFs}
\end{figure}

\begin{figure}[H]
	\centering
	\includegraphics[width=0.8\textwidth]{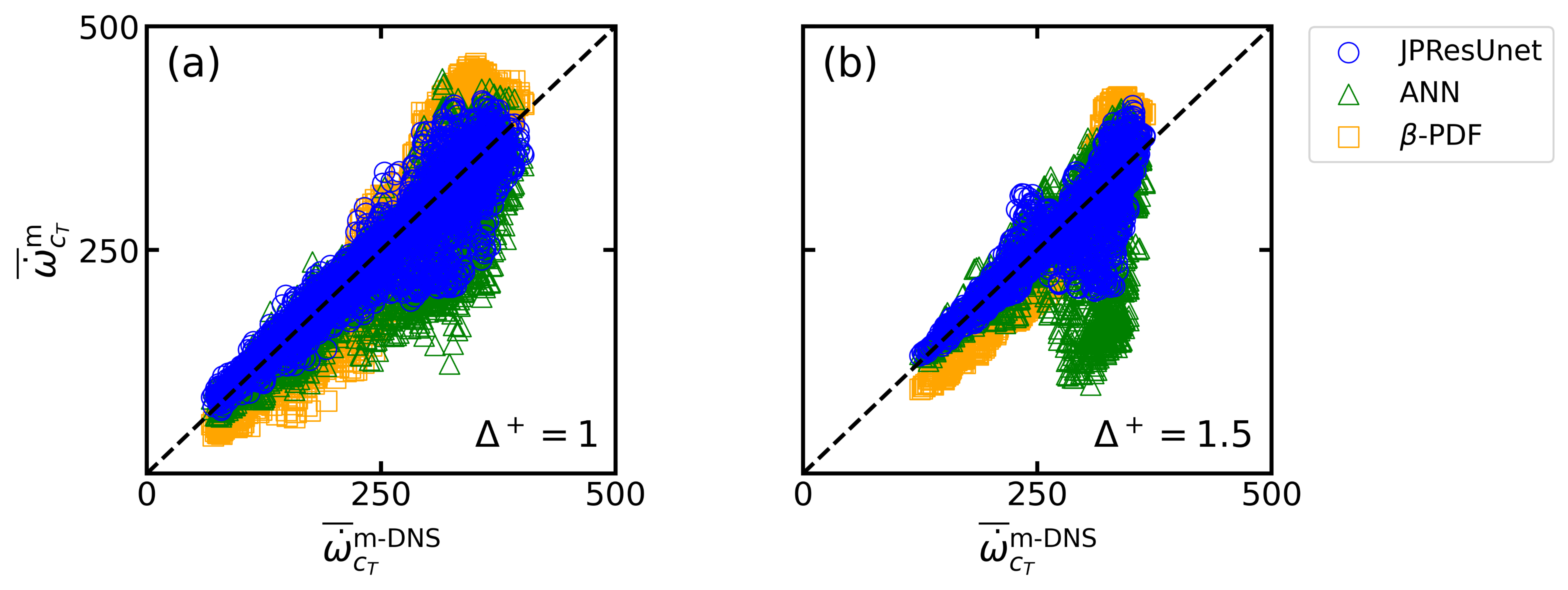}
	\caption{\footnotesize Scatter plot of the progress variable source term modelled by joint PDFs from different models and from the DNS for the case BZ1, using the Gaussian filter with different filter widths.}
	\label{fig:out-sample_BZ1_gaussian_reaction_rate}
\end{figure}

\end{document}